\documentclass[aps,prd,superscriptaddress,10pt,showpacs,notitlepage,  
nofootinbib
]{revtex4-1}
\usepackage{bm}
\usepackage{amsmath,amssymb,amsthm}
\usepackage{latexsym,graphicx,color,subfigure}
\usepackage{enumerate}
\usepackage{amssymb}
\usepackage{hyperref}
\usepackage{mathtools}

\usepackage{graphicx}
\graphicspath{ {images/} }

\usepackage{color}

\newcommand{\be}{\begin{equation}}
\newcommand{\ee}{\end{equation}} 
\newcommand{\beq}{\begin{eqnarray}}
\newcommand{\eeq}{\end{eqnarray}}

\newcommand{\bea}{\begin{eqnarray}}
\newcommand{\eea}{\end{eqnarray}}

\renewcommand{\vec}[1]{\boldsymbol{#1}}
\newcommand\cev[1]{\overleftarrow{#1}}

\allowdisplaybreaks[3]
\def\simge{\mathrel{
   \rlap{\raise 0.511ex \hbox{$>$}}{\lower 0.511ex \hbox{$\sim$}}}}
\def\simle{\mathrel{
   \rlap{\raise 0.511ex \hbox{$<$}}{\lower 0.511ex \hbox{$\sim$}}}}
\def\bigs{\mathrel{
   \rlap{\raise 0.531ex \hbox{$>$}}{\lower 0.531ex \hbox{$<$}}}}

\usepackage[normalem]{ulem}

\renewcommand\sout{\bgroup \color{blue} \ULdepth=-.5ex \ULset}

\begin{document} 
\title{Fate of the charm baryon $\Lambda_{c}$ in cold and hot nuclear matter}
\author{Shigehiro Yasui}
\email{yasuis@keio.jp}
\affiliation{Research and Education Center for Natural Sciences,\\ Keio University, Hiyoshi 4-1-1, Yokohama, Kanagawa 223-8521, Japan}
\date{\today}
\begin{abstract}
I discuss the properties of the $\Lambda_{c}$ baryon in nuclear matter at zero or finite temperature.
Starting from the Lagrangian based on the heavy-quark effective theory, I derive the effective Lagrangian for the $\Lambda_{c}$ baryon existing as an impurity particle.
Adopting the one-loop calculation for nucleons, I derive the effective potential as the quantity for measuring the stability of the $\Lambda_{c}$ baryon in nuclear matter.
The parameters in the Lagrangian are fitted to reproduce the scattering length of the nucleon and the $\Lambda_{c}$ baryon estimated in the lattice QCD simulations and the chiral extrapolations.
I present that the $\Lambda_{c}$ baryon is bound in nuclei with the binding energy of about 20 MeV at normal nuclear-matter density.
I discuss the case that the $\Lambda_{c}$ baryon moves with a constant velocity.
I also discuss an increase of the nucleon number density near the $\Lambda_{c}$ baryon in nuclear matter, and show that the $\Lambda_{c}$ baryon is a useful probe to research the nuclear systems at high density.
\end{abstract}
\maketitle


\setlength\arraycolsep{2pt}

\section{Introduction}

Recently there has been general recognition that the extension of flavors is important for uncovering valuable information on the strong interaction.
In fact, many exotic hadrons, whose structures are significantly different from normal hadrons (baryons and mesons), have been found in charm and bottom flavors at experimental facilities~\cite{Brambilla:2010cs,Brambilla:2014jmp,Chen:2016qju,Hosaka:2016pey,Chen:2016spr,Guo:2017jvc,Ali:2017jda}.
As one of the next problems, the extension of flavors to heavier flavors in nuclear systems is an interesting subject.
Charm (bottom) nuclei contain heavy flavors (charm and bottom) as impurity particles (see, e.g., Refs.~\cite{Hosaka:2016ypm,Krein:2017usp} for a review).
They are qualitatively different from hypernuclei in strangeness, because the masses of charm (bottom) hadrons are heavier than the low-energy scales in QCD.
When a heavy quark exists in the system, one can introduce a new symmetry: the heavy-quark spin symmetry~\cite{Isgur:1989vq,Isgur:1989ed,Isgur:1991wq} (see Ref.~\cite{Manohar:2000dt} for textbook).
This is the symmetry that the heavy-quark spin is decoupled from the light component (light quarks and gluons) in the system in the heavy-quark mass limit.
The heavy-quark spin symmetry provides new pattern in spectroscopy (masses and decays) and reaction, and now it is widely used also in the research of the exotic hadrons with charm and bottom, such as $XYZ$ and $P_{c}$ (see, e.g., Refs.~\cite{Brambilla:2014jmp,Hosaka:2016pey,Chen:2016qju}).
Because the heavy quark symmetry should hold in any hadron systems, one can expect to apply the heavy quark symmetry to a heavy hadron in nuclear matter, which may serve as a novel probe for nuclear systems in a manner different from a strangeness hadron (see, e.g., Refs.~\cite{Hosaka:2016ypm,Krein:2017usp} for a review and the references therein).

I consider a charm baryon to be an impurity particle in nuclear matter.
I focus on a $\Lambda_{c}$ baryon as the simple state in charm nuclei.
The quark content in the $\Lambda_{c}$ baryon is up, down, charm ($udc$), in which the $u$ and $d$ quarks exist as the diquark ($ud$) with an attractive interaction~\cite{Jaffe:1976ig,Jaffe:1976ih,Anselmino:1992vg}.
The $ud$ diquark is also relevant to the color superconductivity in quark matter at high density (see Refs.~\cite{Alford:2007xm,Fukushima:2010bq,Fukushima:2013rx} for a review).
Thus, to study the $ud$ diquark in nuclear matter can be regarded as a first step toward the research of the high-density state.
The same discussion can be applied to a bottom baryon, $\Lambda_{b}$, with better accuracy due to the heavier mass of the bottom quark.

One of the most basic properties about the $\Lambda_{c}$ baryon in nuclear matter is provided by the interaction between a $\Lambda_{c}$ baryon and a nucleon ($N$).
The study of the $\Lambda_{c}N$ interaction dates back to the late 1970s, around the time when the meson-exchange potential was adopted for the $\Lambda_{c}N$ interaction~\cite{Iwao:1976yi}, and the possibility of the $\Lambda_{c}$ bound in atomic nuclei was explored~\cite{Dover:1977jw,Gatto:1978ka,Gibson:1983zw}.
Along the development in the theory of hypernuclei, SU(4) flavor symmetry was considered to be a simple extension of flavor from up, down, strangeness to including charm~\cite{Bando:1981ti,Bando:1983yt,Bando:1985up}.
In those models, the meson-exchange potential of the $\Lambda_{c}N$ interaction was provided as an analogy to the phenomenological nucleon-nucleon and hyperon-nucleon potentials.
Later, the interaction between a $\Lambda_{c}$ baryon and a nucleon was analyzed in terms of the heavy-quark spin symmetry without using SU(4) flavor symmetry~\cite{Liu:2011xc,Meguro:2011nr}.
The possibility of the existence of $\Lambda_{c}N$  and $\Lambda_{c}NN$ bound and/or resonant states was studied in detail~\cite{Garcilazo:2015qha,Maeda:2015hxa,Maeda:2018xcl}, while they were not found in other theoretical studies~\cite{Huang:2013zva,Gal:2014jza} (see also Refs.~\cite{Meguro:2011nr,Li:2012bt,Chen:2013sba,Huang:2013rla,Vijande:2016nzk,Meng:2017fwb,Meng:2017udf}).
Recently, the $\Lambda_{c}N$ potential has been calculated by the lattice QCD simulations.
The results obtained by Miyamoto {\it et al.} indicated that the $\Lambda_{c}N$ interaction is attractive in both $^{1}S_{0}$ and $^{3}S_{1}$ channels and that the difference in the potentials in the two channels is small~\cite{Miyamoto:2017tjs}.
The latter property is in good agreement with the expectations from the heavy-quark spin symmetry.
Nevertheless, the attraction is not sufficiently strong to form $\Lambda_{c}N$ bound states.

Given an attraction between a $\Lambda_{c}$ baryon and a nucleon, there can exist a charm nucleus in which the $\Lambda_{c}$ baryon is bound as the ground state in the strong interaction, as long as the baryon number is sufficiently large.
For simplicity, one may consider the nuclear matter to be an ideal case in which the surface effect can be ignored.
This situation can be realized approximately in the inside of atomic nuclei with large baryon numbers.
In the quark-meson coupling model, it was considered that the change of quark masses at finite density is caused by the partial restoration of the broken chiral symmetry, and it was obtained that the binding energy is around the order of hundred MeV~\cite{Tsushima:2002cc,Tsushima:2002ua,Tsushima:2002sm,Tsushima:2003dd,Tan:2004mt,Tsushima:2018goq}.
The calculation from the QCD sum rules, which is the method directly based on QCD, gave an attraction for the $\Lambda_{c}$ baryon with the binding energy about 20 MeV in nuclear matter~\cite{Ohtani:2017wdc}.
However, it should be kept in mind that there are several studies in the QCD sum rules which rule out the possibility of the $\Lambda_{c}$ baryon bound in nuclear matter, while the $\Sigma_{c}$ baryon is bound~\cite{Wang:2011yj,Azizi:2016dmr} and the $\Sigma_{c}^{\ast}$ baryon also~\cite{Azizi:2018dtb}.

The purpose of the present study is to research the stability of a $\Lambda_{c}$ baryon in nuclear matter in terms of the heavy-quark spin symmetry.
I consider the zero-range interaction between a $\Lambda_{c}$ baryon and a nucleon, and evaluate the stability of the system in the presence of the $\Lambda_{c}$ baryon.
The values of the coupling constant are estimated with a reference to the $\Lambda_{c}N$ potential in the lattice QCD simulations~\cite{Miyamoto:2017tjs}.
However, in Ref.~\cite{Miyamoto:2017tjs}, a heavier pion (whose mass was larger than 410 MeV) was used to perform the calculation, and hence their potential can be different from the realistic one.
In order to carry out a proper evaluation regarding the effective potential, I use the result which was obtained by the chiral extrapolation based on the lattice QCD simulations.
In the work by Haidenbauer and Krein~\cite{Haidenbauer:2017dua}, they estimated the values of the scattering length and the effective range at the real pion mass.
I will use those values in order to constrain the possible range of the parameters.
Under this setup, I will estimate the effective potential in the presence of the $\Lambda_{c}$ baryon in nuclear matter with various temperatures and nucleon densities and will discuss the stability of the $\Lambda_{c}$ baryon in nuclear matter.
I will also discuss the change of the nucleon number density near the $\Lambda_{c}$ baryon and will demonstrate that the $\Lambda_{c}$ baryon is a useful probe to research the higher-density state in nuclear matter.

The article is organized as it follows.
In Sec.~\ref{sec:Lagrangian}, I introduce the interaction Lagrangian for a nucleon and a $\Lambda_{c}$ baryon, and obtain the effective Lagrangian by assuming that the $\Lambda_{c}$ baryon is at rest in nuclear matter.
In Sec.~\ref{sec:eff_pot}, I derive the effective potential in the presence of the $\Lambda_{c}$ baryon in nuclear matter, and also derive the equation expressing the change of nucleon number density near the $\Lambda_{c}$ baryon.
Under this setup, I show the numerical results in Sec.~\ref{sec:numerical_results} and conduct in-depth analyses of the numerical results in Sec.~\ref{sec:discussions}.
The final section is devoted to the conclusion.

\section{Lagrangian based on heavy-quark spin symmetry}
\label{sec:Lagrangian}

\subsection{Effective Lagrangian}

I consider the interaction Lagrangian for a nucleon and a $\Lambda_{c}$ baryon.
I follow the description based on the heavy-quark spin symmetry by supposing that the mass of the $\Lambda_{c}$ baryon, $M=2.286$ GeV, is sufficiently massive in comparison to the typical energy scales in the low energy QCD (a few hundreds of MeV)~\cite{Neubert:1993mb,Casalbuoni:1996pg,Manohar:2000dt}.
I separate the four-momentum of the $\Lambda_{c}$ baryon $p^{\mu}$ as $p^{\mu}=Mv^{\mu}+k^{\mu}$ with $v^{\mu}$ the four-velocity $v^{\mu}=(v^{0},\vec{v})$ ($v^{0}>0$ and $v^{\mu}v_{\mu}=1$) and the residual momentum $k^{\mu}$.
The term $Mv^{\mu}$ indicates the on-mass-shell part, and the term $k^{\mu}$ indicates the off-mass-shell part.
It is supposed that the latter is a small quantity relevant to the low energy QCD, and it is smaller than the mass of the $\Lambda_{c}$ baryon: $k^{\mu} \ll M$.
In the present system, the typical scales of $k^{\mu}$ are the Fermi energy for $k^{0}$ and the Fermi momentum for $k^{1}$, $k^{2}$, and $k^{3}$, and hence $k^{\mu}/M$ should be regarded as a small number so that the expansion in terms of $1/M$ should be valid.
In the framework of the heavy baryon effective theory, instead of the original field of the $\Lambda_{c}$ baryon $\Psi(x)$, I introduce the effective field for the $\Lambda_{c}$ baryon defined by
\begin{eqnarray}
   \Psi_{v}(x) = \frac{1+v\hspace{-0.5em}/}{2} e^{iMv\cdot x} \Psi(x),
\label{eq:Psiv_def}
\end{eqnarray}
with the four-dimensional time and space coordinate $x^{\mu}=(t,\vec{x})$.
In this definition, the $\Lambda_{c}$ baryon is at rest in the coordinate frame moving with the four-velocity $v^{\mu}$ ($v$-frame).
In Eq.~\eqref{eq:Psiv_def}, $(1+v\hspace{-0.5em}/)/2$ is the projection operator to pickup the positive energy state in the $v$-frame, and $e^{-iMv\cdot x}$ represents the on-mass-shell component in $\Psi(x)$.
Thus, $\Psi_{v}(x)$ deals with the off-mass-shell (virtual) component with positive energy component, in which the on-mass-shell component ($e^{-iMv\cdot x}$) is subtracted from $\Psi(x)$.
In the following most cases, I assume the static four-velocity $v^{\mu}=(1,\vec{0})$, i.e., that the $\Lambda_{c}$ baryon is at rest in nuclear matter.

In the relativistic formalism for the nucleon field $\psi$,
considering all the possible combinations of the interaction terms in the $S$-wave,
one obtains the general form of the interaction Lagrangian up to ${\cal O}(1/M)$ given by
\begin{eqnarray}
{\cal L}^{\mathrm{rel}}_{\mathrm{int}}
&=&
   c_{1} \bar{\psi}\psi \bar{\Psi}_{v} \Psi_{v} + \frac{c'_{1}}{M} \bar{\psi}\psi \bar{\Psi}_{v}\Psi_{v}
+ c_{2} \bar{\psi}\gamma^{\mu}\psi
     \bar{\Psi}_{v} \biggl( v_{\mu} - \frac{i\cev{D}_{\perp\mu}}{2M} + \frac{iD_{\perp\mu}}{2M} \biggr) \Psi_{v}
+ \frac{c'_{2}}{M} \bar{\psi}\gamma^{\mu}\psi v_{\mu} \bar{\Psi}_{v}\Psi_{v}
\nonumber \\
&&
+ \frac{1}{M} \Bigl(
     c_{3} \bar{\psi}\sigma^{\mu\nu}\psi 
   \epsilon_{\mu\nu\rho\sigma}v^{\rho}
    + c_{4} \bar{\psi}\gamma_{\sigma}\gamma_{5}\psi
   \Bigr)
    \bar{\Psi}_{v} S_{v}^{\sigma} \Psi_{v}
+ {\cal O}(1/M^{2}),
\label{eq:Lagrangian_int}
\end{eqnarray}
with unknown coefficients $c_{1}$, $c_{2}$, $c'_{2}$, $c_{3}$, and $c_{4}$ and where $m$ is the nucleon mass.
The term $iD_{\perp}^{\mu} \equiv iD^{\mu} - v^{\mu} v \!\cdot\! iD$ is necessary to achieve the velocity-rearrangement (reparametrization) to take into account the terms at ${\cal O}(1/M)$~\cite{Luke:1992cs,Neubert:1993iv,Neubert:1993zc,Kitazawa:1993bk}.
 $S_{v}^{\mu} \equiv -\frac{1}{2} \gamma_{5} \bigl( \gamma^{\mu}v\hspace{-0.5em}/-v^{\mu} \bigr)$ is the spin operator for the $\Lambda_{c}$ baryon.
The terms containing $S_{v}^{\mu}$ should be the order of ${\cal O}(1/M)$ as it is shown in the above equation.
This order counting stems from the fact that the spin flip of the heavy quark is suppressed by $1/M_{Q}$ ($M_{Q}$ the heavy-quark mass) in the heavy-quark effective theory.
One regards $M_{Q} \simeq M$ because $M_{Q}$ is the dominantly large energy scales in the system.
One remarks the spin symmetry for $\Psi_{v}(x)$, i.e., $\Psi_{v}(x) \rightarrow e^{i\vec{\theta}\cdot\vec{\sigma}/2}\Psi_{v}(x)$, with the Pauli matrices $\vec{\sigma}=(\sigma^{1},\sigma^{2},\sigma^{3})$. This stems from the spin symmetry for the heavy quark in the $\Lambda_{c}$ baryon in the heavy-quark mass limit, because the $\Lambda_{c}$ baryon is composed of the spin-zero $ud$ diquark and a spin-1/2 heavy (charm) quark.
In the nonrelativistic limit, the interaction Lagrangian \eqref{eq:Lagrangian_int} becomes a simpler form.
When one keeps only the leading term in the $1/M$ expansion, one confirms that the remaining interaction terms in the Lagrangian \eqref{eq:Lagrangian_int} turn to be $c_{1} \varphi^{\dag}\varphi \bar{\Psi}_{v} \Psi_{v}$ only.
$\varphi$ is the nonrelativistic nucleon field: $\psi^{t}=(\varphi,0)^{t}$.
As a result, one obtains the nonrelativistic Lagrangian
\begin{eqnarray}
 {\cal L}[\varphi,\Psi_{v}]
= \varphi^{\dag}i\frac{\partial}{\partial t}\varphi
  + \varphi^{\dag} \frac{\vec{\nabla}^{2}}{2m}\varphi
  + \bar{\Psi}_{v}i\frac{\partial}{\partial t}\Psi_{v}
  + c_{1} \varphi^{\dag}\varphi \bar{\Psi}_{v} \Psi_{v},
\label{eq:nonrelativistic_Lagrangian}
\end{eqnarray}
with the coupling constant $c_{1}$.
In this formalism, the mass of the $\Lambda_{c}$ baryon $M$ is absorbed into $e^{iMv\cdot x}$ in Eq.~\eqref{eq:Psiv_def}, and the energy of the system is measured from $M$.
Notice that there is no spatial propagation for the $\Lambda_{c}$ baryon, because only the leading terms in ${\cal O}(1/M)$ are considered.

Now let us consider the solutions of the Lagrangian \eqref{eq:nonrelativistic_Lagrangian}.
Before proceeding the discussion, note that the $\Lambda_{c}$ baryon exists as an impurity particle in nuclear matter.
Thus, it is required to impose the condition for the spatial distribution of the number density of the single $\Lambda_{c}$ baryon. This condition is not included in the Lagrangian \eqref{eq:nonrelativistic_Lagrangian}.
I take the case that the $\Lambda_{c}$ baryon is at the spatial position $\vec{x}=\vec{0}$ (the zero point in space), and consider that the $\Lambda_{c}$ baryon is at rest without moving in spatial directions.
This will be a reasonable situation because the mass of the $\Lambda_{c}$ baryon is supposed to be sufficiently heavy.
Then the constraint condition for the number density of the $\Lambda_{c}$ baryon can be imposed\footnote{Notice $\bar{\Psi}_{v}(x)=\Psi^{\dag}(x)$ in the rest frame.}:
\begin{eqnarray}
 \bar{\Psi}_{v}(x)\Psi_{v}(x)=\delta^{(3)}(\vec{x}),
\label{eq:Lambdac_constraint}
\end{eqnarray}
where $\delta^{(3)}(\vec{x})$ is the three-dimensional $\delta$ function.
The density distribution of the $\Lambda_{c}$ baryon is naturally considered to be point like, because the long wavelength scale is adopted for the low-density nuclear matter. This can be verified reasonably, because it is thought that the Fermi wavelength of the nucleons are larger than the spatial size of the $\Lambda_{c}$ baryon. In this limit, the detailed spatial structure of the density distribution should be smeared out, and hence it can be expressed by the $\delta$ function as shown in Eq.~\eqref{eq:Lambdac_constraint}. 
It is possible to extend the present formalism to include the extended distribution for the $\Lambda_{c}$ baryon if necessary.

In order to find a solution for the Lagrangian \eqref{eq:nonrelativistic_Lagrangian} with the condition \eqref{eq:Lambdac_constraint},
one further transforms the Lagrangian \eqref{eq:nonrelativistic_Lagrangian} to a more tractable form.
For this purpose, one considers the generating functional for the Lagrangian \eqref{eq:nonrelativistic_Lagrangian},
\begin{eqnarray}
 Z
 =
 {\cal N}
 \int {\cal D}\psi{\cal D}\bar{\psi} {\cal D}\Psi_{v} {\cal D}\bar{\Psi}_{v}
 \prod_{t,\vec{x}} 
 \delta\bigl(\bar{\Psi}_{v}(x)\Psi_{v}(x)-\delta^{(3)}(\vec{x})\bigr)
 \,
 \exp
 \biggl(
    i\int \mathrm{d}^{4}x \, {\cal L}[\varphi,\Psi_{v}]
 \biggr),
\end{eqnarray}
where ${\cal N}$ is an overall factor irrelevant to the dynamics.
One notices that the constraint condition for the $\Lambda_{c}$ baryon in Eq.~\eqref{eq:Lambdac_constraint}  is accounted for the $\delta$ function in $Z$,
where it is supposed to hold in all times and positions as denoted by $\prod_{t,\vec{x}}$.
At first sight, it might still seem difficult to perform exactly the path integral for $\Psi_{v}$ and $\bar{\Psi}_{v}$.
However, this can be easily resolved by introducing the auxiliary field $\lambda(x)$ (real scalar field) as
\begin{eqnarray}
 \prod_{t,\vec{x}} 
 \delta\bigl(\bar{\Psi}_{v}(x)\Psi_{v}(x)-\delta^{(3)}(\vec{x})\bigr)
=
{\cal N}'
\int {\cal D}\lambda
\exp
\biggl(
-i\int \mathrm{d}^{4}x \, \lambda(x)
\Bigl( \bar{\Psi}_{v}(x)\Psi_{v}(x)-\delta^{(3)}(\vec{x}) \Bigr)
\biggr),
 \label{eq:lambda_definition}
\end{eqnarray}
with an overall factor ${\cal N}'$.
The method of introducing the auxiliary field for treating the constraint condition have been used in the impurity particle systems in the condensed-matter physics~\cite{Newns:1987} (see also Ref.~\cite{Hewson}).
Then, one rewrites the generating functional $Z$ as
\begin{eqnarray}
 Z_{\lambda}
 = 
 {\cal N}''
 \int {\cal D}\psi{\cal D}\bar{\psi} {\cal D}\Psi_{v} {\cal D}\bar{\Psi}_{v} {\cal D}\lambda  \,
 \exp
 \biggl(
    i\int \mathrm{d}^{4}x \, {\cal L}[\psi,\Psi_{v},\lambda]
 \biggr),
\end{eqnarray}
with an overall factor ${\cal N}''={\cal N}{\cal N}'$,
where the new Lagrangian is defined by
\begin{eqnarray}
 {\cal L}[\varphi,\Psi_{v},\lambda]
&=& {\cal L}[\varphi,\Psi_{v}]
 - \lambda(x) \Bigl( \bar{\Psi}_{v}(x)\Psi_{v}(x)-\delta^{(3)}(\vec{x}) \Bigr)
\nonumber \\
&=& \varphi^{\dag}i\frac{\partial}{\partial t}\varphi
  + \varphi^{\dag} \frac{\vec{\nabla}^{2}}{2m}\varphi
  + \bar{\Psi}_{v}i\frac{\partial}{\partial t}\Psi_{v}
  + c_{1} \varphi^{\dag}\varphi \bar{\Psi}_{v} \Psi_{v}
  - \lambda \Bigl( \bar{\Psi}_{v}\Psi_{v}-\delta^{(3)}(\vec{x}) \Bigr).
\end{eqnarray}
Furthermore, one replaces the auxiliary field as
\begin{eqnarray}
 \lambda \rightarrow \lambda + c_{1} \varphi^{\dag}\varphi,
 \label{eq:lambda_shift}
\end{eqnarray}
which does not change the dynamics essentially.
As a result, one obtains the new form of the Lagrangian,
\begin{eqnarray}
 {\cal L}[\varphi,\Psi_{v},\lambda] = {\cal L}[\varphi] + {\cal L}[\Psi_{v},\lambda],
\label{eq:L_L1_L2}
\end{eqnarray}
where ${\cal L}[\varphi]$ and ${\cal L}[\Psi_{v},\lambda]$ are defined by
\begin{eqnarray}
   {\cal L}[\varphi]
=
     \varphi^{\dag}i\frac{\partial}{\partial t}\varphi
  + \varphi^{\dag} \frac{\vec{\nabla}^{2}}{2m}\varphi
  + c_{1} \varphi^{\dag}\varphi \, \delta^{(3)}(\vec{x})
\label{eq:eff_Lagrangian_N}
\end{eqnarray}
and
\begin{eqnarray}
{\cal L}[\Psi_{v},\lambda]
=
  \bar{\Psi}_{v}i\frac{\partial}{\partial t}\Psi_{v}
  - \lambda \bar{\Psi}_{v}\Psi_{v} + \lambda \, \delta^{(3)}(\vec{x}),
\label{eq:eff_Lagrangian_Lambdac_lambda}
\end{eqnarray}
respectively.
It is important to note that, in the separation of the terms in Eq.~\eqref{eq:L_L1_L2}, the nucleon ($\varphi$) is decoupled from the $\Lambda_{c}$ baryon ($\Psi_{v}$) and from the auxiliary field ($\lambda$).
The dynamics of the nucleon is irrelevant to $\Psi_{v}$, $\bar{\Psi}_{v}$, and $\lambda$,
and the Lagrangian relevant to the nucleon dynamics is provided only by ${\cal L}[\varphi]$ in Eq.~\eqref{eq:eff_Lagrangian_N}.
One notes that the path integral about ${\cal L}[\Psi_{v},\lambda]$ does not provide any information about the nuclear medium, and thus it is irrelevant in the present purpose.
Therefore, ${\cal L}[\varphi]$ is regarded as the basic effective Lagrangian in the following discussions.
One can check that Eq.~\eqref{eq:eff_Lagrangian_N} is also obtained by substituting Eq.~\eqref{eq:Lambdac_constraint} into Eq.~\eqref{eq:nonrelativistic_Lagrangian} in a straightforward manner.
The advantage of introducing the auxiliary field $\lambda$ is the general applicability for higher-order terms.
It is commented that the Lagrangian \eqref{eq:eff_Lagrangian_N} is essentially the same as the Clogston model which has been used for impurity systems in the condensed-matter physics~\cite{PhysRev.125.439}.
It is also commented that a pion is not included in the Lagrangian.
This is because there is no interaction between a pion and a $\Lambda_{c}$ baryon due to the zero isospin of the $\Lambda_{c}$ baryon.
Finally it is mentioned that $\Sigma_{c}$ and $\Sigma_{c}^{\ast}$ baryons are not taken into account in the present study, because the mass splitting between the $\Sigma_{c}$ ($\Sigma_{c}^{\ast}$) baryon and the $\Lambda_{c}$ baryon is too large in the relevant energy scales in the present temperature and Fermi energy.

\subsection{$T$ matrix in vacuum}

For the effective Lagrangian \eqref{eq:eff_Lagrangian_N}, one constrains the value of the coupling constant $c_{1}$.
To estimate it, one utilizes the result by the lattice QCD simulation.
Recently, Miyamoto {\it et al.} gave the potential between a nucleon and a $\Lambda_{c}$ baryon by using the HAL-QCD method~\cite{Miyamoto:2017tjs}.
The obtained potentials are attractive at long distances and repulsive at short distances.
Because the pion masses used in their simulations are not so close to the real value, further analysis is needed.
Based on the result by Ref.~\cite{Miyamoto:2017tjs}, Haidenbauer and Krein adopted the chiral perturbation theory for the $\Lambda_{c}N$ interaction, and they obtained the scattering length and the effective range at the real pion mass~\cite{Haidenbauer:2017dua}.
I use the value of the scattering length in Ref.~\cite{Haidenbauer:2017dua} in order to constrain the range of values of $c_{1}$ in Eq.~\eqref{eq:eff_Lagrangian_N}.

I consider the scattering process of the nucleon scattered on a $\Lambda_{c}$ baryon in vacuum.
I suppose that the nucleon has the energy and momentum, $(\omega_{\vec{p}},\vec{p})$ with $\omega_{\vec{p}}=\vec{p}^{2}/(2m)$ in the initial state and $(\omega_{\vec{p}'},\vec{p}')$ with $\omega_{\vec{p}'}=\vec{p}^{\prime2}/(2m)$ in the final state.
Then, starting from the Lagrangian \eqref{eq:eff_Lagrangian_N} and taking the multiple scatterings by the $\Lambda_{c}$ baryon into account, one finds the $T$ matrix given by
\begin{eqnarray}
 iT(\omega_{\vec{p}}) \, 2\pi \delta(\omega_{\vec{p}}-\omega_{\vec{p}'}) \mathbf{1}
=
 \cfrac{ic_{1}}
 {1+
 c_{1}
 \displaystyle \int \frac{\mathrm{d}^{3}\vec{k}}{(2\pi)^{3}}
 \frac{1}{\omega_{\vec{p}}-\frac{\vec{k}^{2}}{2m}+i\varepsilon}
 }
 2\pi \delta(\omega_{\vec{p}}-\omega_{\vec{p}'}) \mathbf{1},
\label{eq:sum_T_matrix}
\end{eqnarray}
which is a sum of an infinite series of $c_{1}$ and $\vec{1}$ is a unit matrix in the spin and isospin space.
I introduce a small and positive quantity $\varepsilon$.
The momentum integral by the three-dimensional momentum $\vec{k}$ in the denominator includes the off-mass-shell motion of the nucleons in the multiple scatterings.
For the regularization of the momentum integral, I introduce the sharp cutoff parameter $\Lambda$ and restrict the integral region for $k=|\vec{k}|$ as $k \in [0,\Lambda]$.
This regularization procedure is adopted because the theory with the four-fermion interaction is not a renormalizable one. Physically, the inverse of $\Lambda$ can be regarded as the size of the nucleon or the $\Lambda_{c}$ baryon. One will see later that the binding energy of the $\Lambda_{c}$ baryon in nuclear matter has no strong dependence on the choice of $\Lambda$.
The $T$ matrix in Eq.~\eqref{eq:sum_T_matrix} can be expressed in terms of the phase shift $\delta(\omega_{\vec{p}})$ as
\begin{eqnarray}
 \frac{m}{2\pi} T(\omega_{\vec{p}}) = \frac{e^{2i\delta(\omega_{\vec{p}})}-1}{2i|\vec{p}|}.
\end{eqnarray}
Then, the scattering length is obtained as
\begin{eqnarray}
 a
&=&
  \lim_{p\rightarrow0} \frac{1}{p} \tan \delta(\omega_{\vec{p}})
=
  \frac{\pi c_{1}m}{2\pi^{2}-2c_{1}\Lambda m},
\end{eqnarray}
with $p=|\vec{p}|$.
I will use the last equation in order to constrain the value ranges of $\Lambda$ and $c_{1}$ for the given scattering length $a$.

\section{Effective potential of $\Lambda_{c}$ baryon in nuclear matter}
\label{sec:eff_pot}

\subsection{Effective potential at rest frame}

\begin{figure}[t]
\begin{center}
\vspace{0em}
\includegraphics[scale=0.45]{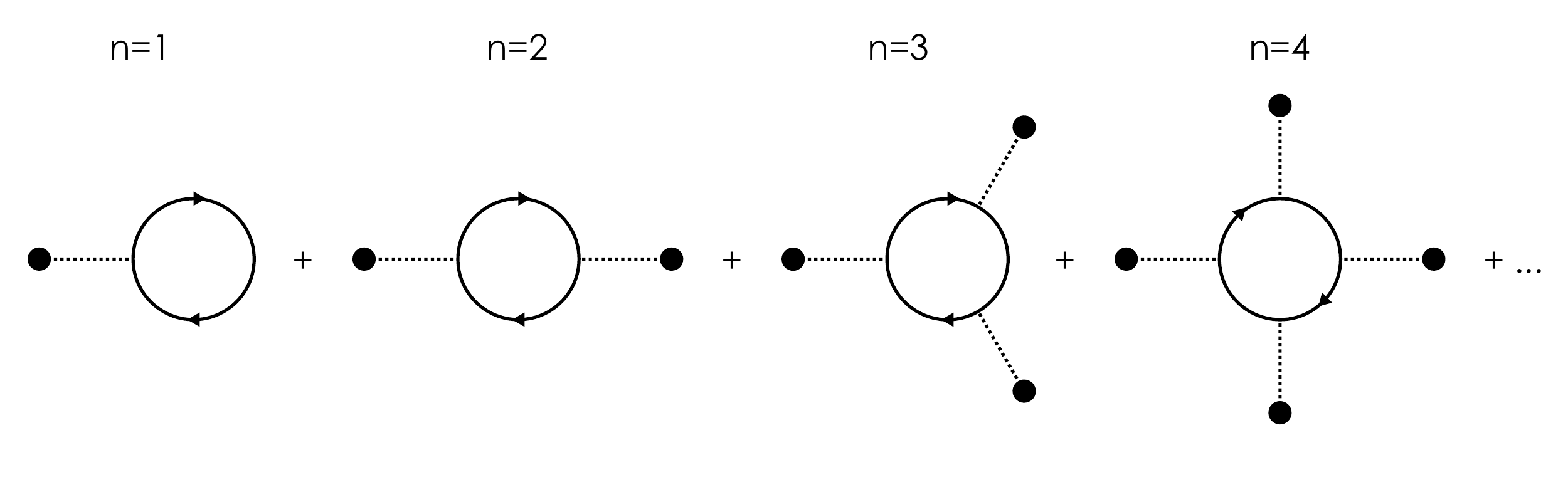}
\vspace{0em}
\caption{The sum of the ring diagrams for calculating the effective potential. The solid lines are the nucleon propagator, and the blobs are the delta-function in Eq.~\eqref{eq:eff_Lagrangian_N}. The dotted lines indicate the  zero-range interaction.}
\label{fig:Fig_effective_potential_classical}
\end{center}
\end{figure}

I consider the energy gain of the system in the presence of a $\Lambda_{c}$ baryon in nuclear matter.
Adopting the Lagrangian \eqref{eq:eff_Lagrangian_N}, one obtains the effective potential of the system
\begin{eqnarray}
-iV^{(0)}
=
 4
 \int \frac{\mathrm{d}p_{0}}{2\pi}
\log \Biggl( 1 + c_{1} \int \dfrac{\mathrm{d}^{3}\vec{p}}{(2\pi)^{3}} \dfrac{1}{p_{0}-\omega_{\vec{p}}
 } \Biggr)
\label{eq:eff_pot}
\end{eqnarray}
in the one-loop calculation for nucleons with $\omega_{\vec{p}}=\vec{p}^{2}/(2m)$.
The momentum integral is performed for the range of $|\vec{p}| \in [0,\Lambda]$.
The relevant diagrams are shown in Fig.~\ref{fig:Fig_effective_potential_classical}.
The coefficient is the number of degeneracy by spin and isospin of a nucleons ($2\times2=4$).
One notices that $-iV^{(0)}$ in Eq.~\eqref{eq:eff_pot} gives an exact solution for Eq.~\eqref{eq:eff_Lagrangian_N}.
The effective potential in Eq.~\eqref{eq:eff_pot} supplies only the energy difference between the case situation that the $\Lambda_{c}$ baryon is present in nuclear matter and the situation that it is absent.
Thus, the effective potential $V^{(0)}$ is a useful quantity measuring the binding energy of the $\Lambda_{c}$ baryon in nuclear matter.
Considering the nuclear matter at finite temperature and density, I use the formula for the Matsubara sum: The $p_{0}$ integral is replaced as
\begin{eqnarray}
 \int \frac{\mathrm{d}p_{0}}{2\pi} f(p_{0})
&\rightarrow&
 \frac{i}{\beta} \sum_{n \in \mathbb{Z}} f(i\omega_{n}+\mu)
=
  \frac{-1}{2\pi} 
  \int_{-\infty}^{\infty}
  \mathrm{d}p_{0}
  \bigl( f(p_{0}+i\varepsilon) - f(p_{0}-i\varepsilon) \bigr)
  \frac{1}{e^{\beta (p_{0}-\mu)}+1},
\label{eq:finite_temperature_2}
\end{eqnarray}
for an analytic function $f(p_{0})$,
where the last equation is presented for $f(p_{0})$ which has a branch cut on the real axis such as a logarithmic function in Eq.~\eqref{eq:eff_pot}.
I define the chemical potential for nucleons $\mu$ and the Matsubara frequencies $\omega_{n}=(2n+1)\pi / \beta$ ($n \in \mathbb{Z}$) with the inverse temperature $\beta=1/T$.
Then, calculating Eq.~\eqref{eq:eff_pot} with the procedure of Eq.~\eqref{eq:finite_temperature_2}, one finds
\begin{eqnarray}
 V^{(0)}(T,\mu)
=
 - \frac{4}{\pi}  \int_{-\infty}^{\infty} \mathrm{d}p_{0} \,
    \arctan
    \Biggl(
       \frac{\pi \, c_{1} \rho(p_{0})}{1+c_{1}F(p_{0})}
    \Biggr)
 \frac{1}{e^{\beta (p_{0}-\mu)}+1},
\label{eq:effective_potential_Lambdac_formfactor}
\end{eqnarray}
with the definition
\begin{eqnarray}
   F(p_{0})
&=&
   \cfrac{\sqrt{2}m^{3/2}}{2\pi^{2}} \, \mathrm{P} \displaystyle \int_{0}^{\frac{\Lambda^{2}}{2m}} \mathrm{d}\omega \frac{\sqrt{\omega}}{p_{0}-\omega},
\label{eq:def_F} \\
   \rho(p_{0})
&=&
   \cfrac{\sqrt{2}m^{3/2}}{2\pi^{2}} \sqrt{p_{0}},
\label{eq:def_rho}
\end{eqnarray}
in which $\mathrm{P}$ stands for the principal-value integral.
$\rho(p_{0})$ indicates the density-of-state at the energy $p_{0}$ for a nucleon without a factor of degeneracy by spin and isospin.

\subsection{Effective potential in moving frame and effective mass}

Next I consider that the $\Lambda_{c}$ baryon is moving with a constant three-dimensional velocity $\vec{u}$ in nuclear matter.
One notices that $\vec{u}$ is related to the spatial component of the four-velocity:
$\vec{v}=\vec{u}/\sqrt{1-|\vec{u}|^{2}}$.
For the small velocity, having $\vec{v} \simeq \vec{u}$,
I replace $\delta^{(3)}(\vec{x})$ with $\delta^{(3)}(\vec{x}-\vec{u}t)$ in Eq.~\eqref{eq:Lambdac_constraint}.
Noting that the zero point of the time $t$ can be chosen arbitrarily, I assume that the $\Lambda_{c}$ baryon exists at $\vec{x}=\vec{0}$ at $t=0$.
Repeating the previous discussions, one finds that the introduction of $\vec{u}$ leads to the change in Eq.~\eqref{eq:def_F}:
\begin{eqnarray}
   \int \frac{\mathrm{d}^{3}\vec{p}}{(2\pi)^{3}}
   \frac{1}{p_{0}-\omega_{\vec{p}}
   }
&\rightarrow&
   \int \frac{\mathrm{d}^{3}\vec{p}}{(2\pi)^{3}}
   \frac{1}{p_{0}+\dfrac{1}{2}mu^{2}-\omega_{\vec{p}}
   },
\end{eqnarray}
with $u=|\vec{u}|$.
This procedure is understood by the replacement of the nucleon momentum from $\vec{p}$ to $\vec{p}-m\vec{u}$ in the right-hand side.
One notices that the situation where a $\Lambda_{c}$ baryon moves with velocity $\vec{u}$ in nuclear matter is equivalent to the situation where the $\Lambda_{c}$ baryon is at rest in the nuclear matter and the nuclear matter moves with velocity $-\vec{u}$. 
As a result, one finds that the effective potential in Eq.~\eqref{eq:effective_potential_Lambdac_formfactor} is changed to
\begin{eqnarray}
  V^{(0)}(T,\mu;\vec{u})
=
 -\frac{4}{\pi}
 \int_{-\infty}^{\infty} \mathrm{d}p_{0}
    \arctan
    \Biggl(
       \frac{\pi \, c_{1} \rho(p_{0})}{1+c_{1}F(p_{0})}
    \Biggr)
 \frac{1}{e^{\beta (p_{0}-\tilde{\mu}_{u})}+1},
\label{eq:eff_pot_u_T}
\end{eqnarray}
with the definition
\begin{eqnarray}
   \tilde{\mu}_{u} = \mu-\frac{1}{2}mu^{2},
\end{eqnarray}
where $\tilde{\mu}$ is called the effective chemical potential.
One notices that the effective chemical potential can become negative for the velocity larger than the critical velocity $u_{c}=\sqrt{2\mu/m}$.
Thus, the velocity should be limited in the range of $0 \le u \le u_{c}$.

Assuming a small velocity with $u \ll u_{c}$, one expands the effective potential in Eq.~\eqref{eq:eff_pot_u_T} as
\begin{eqnarray}
   V^{(0)}(T,\mu;\vec{u}) = V^{(0)}(T,\mu) + \frac{1}{2}M^{(0)}(T,\mu) \vec{u}^{2} + {\cal O}(\vec{u}^{4}),
\label{eq:eff_pot_u_expansion}
\end{eqnarray}
with the definition
\begin{eqnarray}
   M^{(0)}(T,\mu) = 2 \frac{\partial V^{(0)}(T,\mu;\vec{u})}{\partial \vec{u}^{2}} \biggr|_{\vec{u}=\vec{0}}.
\label{eq:eff_mass_eff_pot_u_T_def}
\end{eqnarray}
Substituting Eq.~\eqref{eq:eff_pot_u_T} into Eq.~\eqref{eq:eff_mass_eff_pot_u_T_def}, one obtains
\begin{eqnarray}
   M^{(0)}(T,\mu)
=
 \frac{2\beta\,m}{\pi}
 \int_{-\infty}^{\infty} \mathrm{d}p_{0}
 \arctan
    \Biggl(
       \frac{\pi \, c_{1} \rho(p_{0})}{1+c_{1}F(p_{0})}
    \Biggr)
 \frac{e^{\beta(p_{0}-\mu)}}{\bigl(e^{\beta(p_{0}-\mu)}+1\bigr)^{2}}.
\label{eq:eff_mass_eff_pot_u_T}
\end{eqnarray}
One notices that the integrand in the $p_{0}$ integral has a sharp peak around the Fermi surface ($p_{0} \simeq \mu$) at low temperature.
$M^{(0)}(T,\mu)$ is called the effective mass, because it is the quantity relevant to the mass of inertia of the $\Lambda_{c}$ baryon in nuclear matter.
I investigate the details of the physical meaning of $M^{(0)}(T,\mu)$.
When the $\Lambda_{c}$ baryon mass $M$ is recovered in the total energy, the mass of the $\Lambda_{c}$ at rest in nuclear matter can be expressed by
\begin{eqnarray}
   E_{\Lambda_{c}}^{\ast}(T,\mu)
&=&
   M + V^{(0)}(T,\mu).
\label{eq:energy_0}
\end{eqnarray}
For the moving $\Lambda_{c}$ baryon with with three-dimensional velocity $\vec{u}$ in nuclear matter, by using Eq.~\eqref{eq:eff_pot_u_expansion}, one expresses the energy dispersion relation of the $\Lambda_{c}$ baryon as
\begin{eqnarray}
   E_{\Lambda_{c}}^{\ast}(T,\mu;\vec{u})
&=&
   M + \frac{M}{2} \vec{u}^{2} + V^{(0)}(T,\mu;\vec{u}) + {\cal O}(u^{4})
\nonumber \\
&=&
   M+V^{(0)}(T,\mu) + \frac{M+M^{(0)}(T,\mu)}{2} \vec{u}^{2} + {\cal O}(u^{4}),
\label{eq:energy_Yasui}
\end{eqnarray}
where, in the second line, the functions of $V^{(0)}(T,\mu;\vec{u})$ and $M^{(0)}(T,\mu)$ are defined in the above expansion for small $u$.
In this form, one understands clearly that $M+V^{(0)}(T,\mu)$ and $M+M^{(0)}(T,\mu)$ are the quantities with different physical meanings: The former is the energy of the $\Lambda_{c}$ baryon at rest and the latter is the mass of inertia of the $\Lambda_{c}$ baryon moving in nuclear matter.
Also one notices that the nonrelativistic kinetic energy $M\vec{u}^{2}/2$ is added as the kinetic term in the first line, because  the zero point of energy should be shifted from $M$ to $M+M\vec{u}^{2}/2$ at finite velocity.

\subsection{Change of number density of nucleons}

The presence of a $\Lambda_{c}$ baryon in nuclear matter leads to the disturbance of the nucleon number density according to attraction or repulsion of nucleons to the $\Lambda_{c}$ baryon.
For the $\Lambda_{c}$ baryon existing at the position $\vec{x}=\vec{0}$ statically,
one expresses the nucleon number density modified by the $\Lambda_{c}$ baryon as
\begin{eqnarray}
   n_{N}^{\ast}(T,\mu;\vec{x}) = n_{N}(T,\mu) + \delta n_{N}(T,\mu;\vec{x}),
\label{eq:n_density_Lambdac}
\end{eqnarray}
where $n_{N}(T,\mu)$ is the number density of a free nucleon gas in the bulk space without the presence of a $\Lambda_{c}$ baryon, and $\delta n_{N}(T,\mu;\vec{x})$ is its deviation by the effect of a $\Lambda_{c}$ baryon.
Here let us remind that the nucleon number density can be obtained by the loop integral of the nucleons (see, e.g., Ref.~\cite{doi:10.1142/p067}).
For example, the number density of the free nucleon gas in the bulk space is given as
\begin{eqnarray}
 n_{N}(T,\mu)
=
 -4
 \lim_{\vec{y}\rightarrow\vec{x}}
 \int \frac{\mathrm{d}p_{0}}{2\pi} \frac{\mathrm{d}^{3}\vec{p}}{(2\pi)^{3}}
 \frac{i}{p_{0}-\frac{\vec{p}^{2}}{2m}}
 e^{i\vec{p}\cdot(\vec{x}-\vec{y})},
\label{eq:delta_n_bulk_def}
\end{eqnarray}
with $-4$ the coefficient for the fermion loop and the number of degeneracy by spin and isospin.
This turns to be
\begin{eqnarray}
 n_{N}(T,\mu)
=
 4
 \int \frac{\mathrm{d}^{3}\vec{p}}{(2\pi)^{3}}
 \frac{1}{e^{\beta (\omega_{\vec{p}}-\mu)}+1},
\end{eqnarray}
which in fact coincides with the correct result.
One notices that the $p_{0}$ integral in Eq.~\eqref{eq:delta_n_bulk_def} is calculated by following the procedure in Eq.~\eqref{eq:finite_temperature_2}.
It is obvious that there is no position dependence in $n_{N}(T,\mu)$ in the bulk space.
However, this is not the case when there is a $\Lambda_{c}$ baryon as an impurity particle, because the existence of the $\Lambda_{c}$ baryon violates the translational symmetry and the position dependence should appear.
In the presence of a $\Lambda_{c}$ baryon, the deviation of nucleon number density from the one in bulk space is given as
\begin{eqnarray}
  \delta n_{N}(T,\mu;\vec{x})
&=&
 -4
 \lim_{\vec{y}\rightarrow\vec{x}}
 \int \frac{\mathrm{d}p_{0}}{2\pi}
 \int \frac{\mathrm{d}^{3}\vec{p}}{(2\pi)^{3}}
 \frac{i}{p_{0}-\frac{\vec{p}^{2}}{2m}} e^{i\vec{p}\cdot\vec{x}}
 \int \frac{\mathrm{d}^{3}\vec{q}}{(2\pi)^{3}}
 \frac{i}{p_{0}-\frac{\vec{q}^{2}}{2m}} e^{-i\vec{q}\cdot\vec{y}} \,
 iT(p_{0}),
\label{eq:delta_n_x_def}
\end{eqnarray}
at the position $\vec{x}$.
One notices again that $-4$ the coefficient for the fermion loop and the number of degeneracy by spin and isospin, and that the $p_{0}$ integral is calculated by the the procedure in Eq.~\eqref{eq:finite_temperature_2}.
$T(p_{0})$ is the $T$ matrix in Eq.~\eqref{eq:sum_T_matrix}.
For simplicity of the calculation, one considers $T(p_{0}) \simeq c_{1}$ as the lowest order approximation for the small coupling constant.
Thus, adopting the spherical wave expansion for $\vec{x}$, one obtains
\begin{eqnarray}
   \delta n_{N}(T,\mu;\vec{x})
\simeq
 -8c_{1}
 \biggl( \frac{\sqrt{2}m^{3/2}}{2\pi^{2}} \biggr)^{2}
 \int_{0}^{\infty} \mathrm{d}\omega \,
 \frac{\sqrt{\omega} \, j_{0}(\sqrt{2m\,\omega}\,r)}{e^{\beta(p_{0}-\mu)}+1}
 \int_{0}^{\frac{\Lambda^{2}}{2m}} \mathrm{d}\omega' \, 
 \mathrm{P} \, \frac{\sqrt{\omega'}\,j_{0}(\sqrt{2m\,\omega'}\,r)}{\omega-\omega'},
\label{eq:delta_n_r}
\end{eqnarray}
with $r=|\vec{x}|$ the distance from the position of the $\Lambda_{c}$ baryon.
Just on site of the $\Lambda_{c}$ baryon ($\vec{x}=\vec{0}$), one obtains the simple analytic solution as
\begin{eqnarray}
   \delta n_{N}(T,\mu;\vec{0})
\simeq
 -8c_{1}
 \Biggl( \frac{\sqrt{2}m^{3/2}}{2\pi^{2}} \Biggr)^{2}
 \Biggl( \frac{\Lambda^{2}}{2m} -\mu \Biggr)
 \Biggl(
      \sqrt{\frac{\Lambda^{2}\mu}{2m}}
    - \Biggl( \frac{\Lambda^{2}}{2m} +\mu \Biggr)
      \mathrm{arccoth}\Biggl( \sqrt{\frac{\Lambda^{2}}{2m\mu}} \Biggr)
 \Biggr).
\label{eq:delta_n_r0}
\end{eqnarray}

\section{Numerical results}
\label{sec:numerical_results}

\subsection{Parameter sets}

In order to constrain the parameter values of $\Lambda$ and $c_{1}$,  I use the scattering length $a=0.89$ fm for the interaction between a nucleon and a $\Lambda_{c}$ baryon in vacuum as the input~\cite{Haidenbauer:2017dua}.
From Eq.~\eqref{eq:finite_temperature_2} one obtains the several solutions of the parameter sets for $(\Lambda, c_{1})$ as summarized in Table~\ref{table:sharp_cutoff}: (a) (0.3 GeV, 16.2 GeV$^{-2}$), (b) (0.4 GeV, 14.0 GeV$^{-2}$), and (c) (0.5 GeV, 12.4 GeV$^{-2}$).
I choose the range of the cutoff parameter $\Lambda$ to be the order of a few hundred MeV, because its inverse $1/\Lambda$ should be comparable with the spatial size of hadrons. 

\begin{table}[t]
\caption{The parameter sets (a), (b), and (c) for the sharp cutoff parameter $\Lambda$ and the coupling constant $c_{1}$ are shown. For each parameter set, the effective potential $V^{(0)}=V^{(0)}(T,\mu)$, the effective mass $M^{(0)}=M^{(0)}(T,\mu)$ and the nucleon number density $n_{N}^{\ast}(\vec{0})/n_{N}=n_{N}^{\ast}(T,\mu;\vec{0})/n_{N}(T,\mu)$ are shown at $T=0$ MeV and $\mu=38$ MeV ($n_{N}=0.17$ fm$^{-3}$). Notice the value of $n_{N}^{\ast}(\vec{0})/n_{N}$ is obtained in the approximation leaving only the leading order in the expansion for $c_{1}$. For a comparison, the mass shift estimated in the $T\rho$ approximation is shown in the last row in the column of $V^{(0)}$.}
\begin{center}
\renewcommand{\arraystretch}{1.0}
\begin{tabular}{|c|c|c|c|c|c|}
\hline
  \multicolumn{2}{|c|}{parameter set} & (a) & (b) & (c) & \\
\hline
  \multicolumn{2}{|c|}{$\Lambda$ (GeV)} & 0.3 & 0.4 & 0.5 & \\
  \multicolumn{2}{|c|}{$c_{1}$ (GeV$^{-2}$)} & 16.2 & 14.0 & 12.4 & \\
\hline 
   \multicolumn{2}{|c|}{$V^{(0)}$ (MeV)} & -24.3 & -26.7 & -28.1 & -39.4 ($T\rho$) \\
   \multicolumn{2}{|c|}{$M^{(0)}$ (MeV)} & 308 & 382 & 415 & -- \\
   \multicolumn{2}{|c|}{$n_{N}^{\ast}(\vec{0})/n_{N}$ (approx.)} & 1.31 & 1.73 & 1.95 & -- \\
\hline
\end{tabular}
\renewcommand{\arraystretch}{1.0}
\end{center}
\label{table:sharp_cutoff}
\end{table}%

\subsection{Effective potential at rest frame}
\label{sec:result_eff_pot_rest}

In Table~\ref{table:sharp_cutoff}, I show the results for the effective potential, Eq.~\eqref{eq:effective_potential_Lambdac_formfactor}, for the $\Lambda_{c}$ baryon at rest in nuclear matter.
They are the results at zero temperature and at normal nuclear-matter density, $T=0$ MeV and $\mu=38$ MeV ($n_{N}=0.17$ fm$^{-3}$).
The values of the obtained effective potentials are in the range from $-24.3$ MeV to $-28.1$ MeV for the different parameter sets (a), (b), and (c).
It is interesting to compare those values with the $\Lambda_{c}$ mass shift in the $T\rho$ approximation:
\begin{eqnarray}
   \Delta M_{T\rho}^{(0)} = -2\pi \, n_{N} a \lim_{M\rightarrow\infty}\frac{m+M}{m M} = -39.4 \,\, \mathrm{MeV}.
\label{eq:T_rho_approximation}
\end{eqnarray}
I consider the heavy mass limit for the $\Lambda_{c}$ baryon ($M \rightarrow \infty$) in order to be consistent with the leading-order approximation in the $1/M$ expansion, as presented in the Lagrangian \eqref{eq:eff_Lagrangian_N}.
When one keeps the finite value of the $\Lambda_{c}$ mass ($M=2.286$ GeV), one obtains $\Delta M_{\Lambda_{c}} = -2\pi \, n_{N} a (m+M)/(m M)  =-55.6$ MeV, which is larger by about 30 \% than the value in Eq.~\eqref{eq:T_rho_approximation}.
In any case, the values in the $T\rho$ approximation overestimates the value of the effective potential \eqref{eq:effective_potential_Lambdac_formfactor}.
Therefore, one finds it important to include the multiple scatterings in the loop expansion in the loop calculation.
It is interesting that the values of $V^{(0)}$ are consistent with the ones obtained in the QCD sum rule calculations.
Ohtani {\it et al.} gave the mass shift of the $\Lambda_{c}$ baryon by $-20$ MeV at normal nuclear-matter density~\cite{Ohtani:2017wdc}.

\begin{figure}[tb]
\begin{center}
\vspace{0cm}
\includegraphics[scale=0.22]{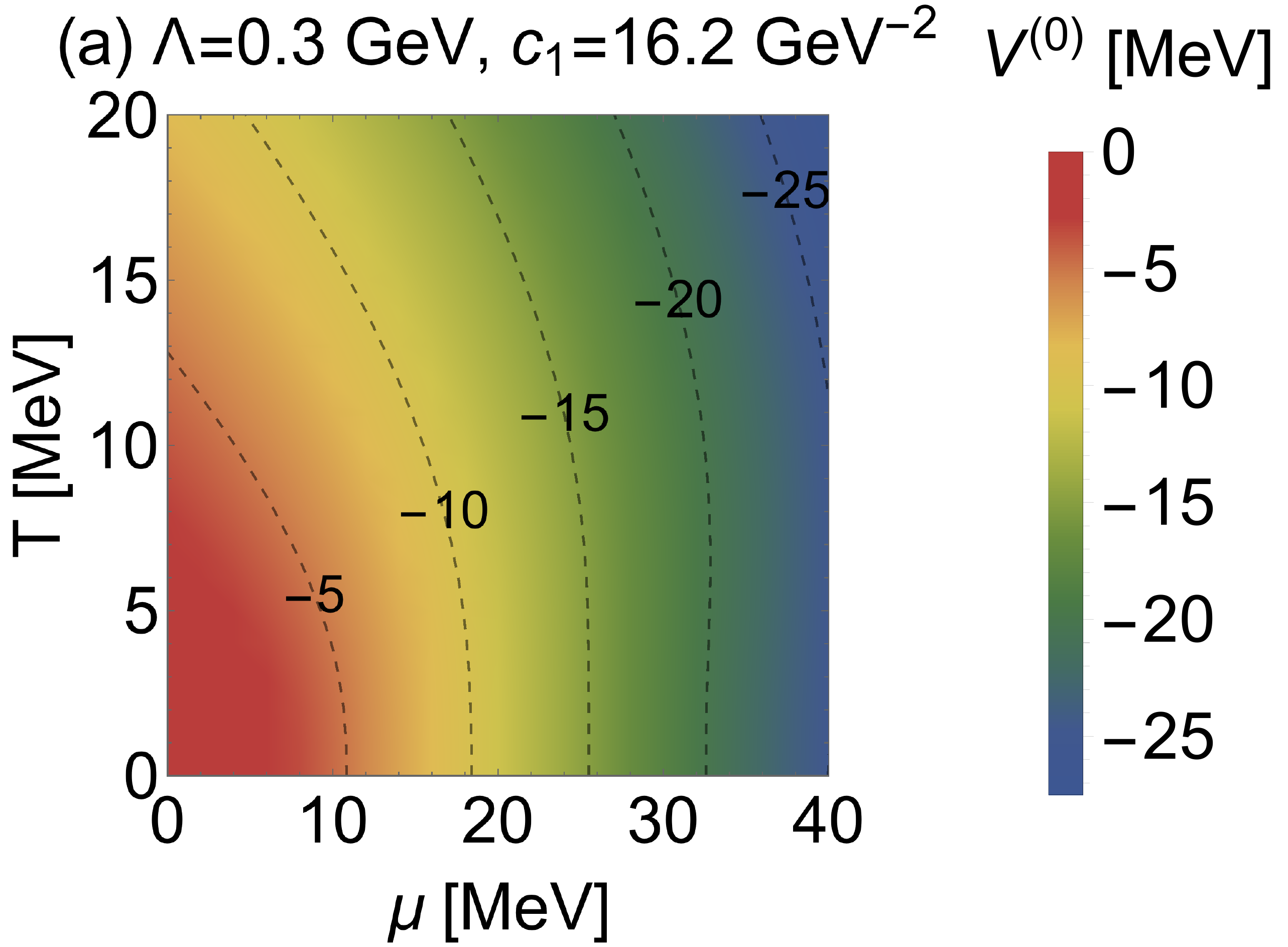}
\includegraphics[scale=0.22]{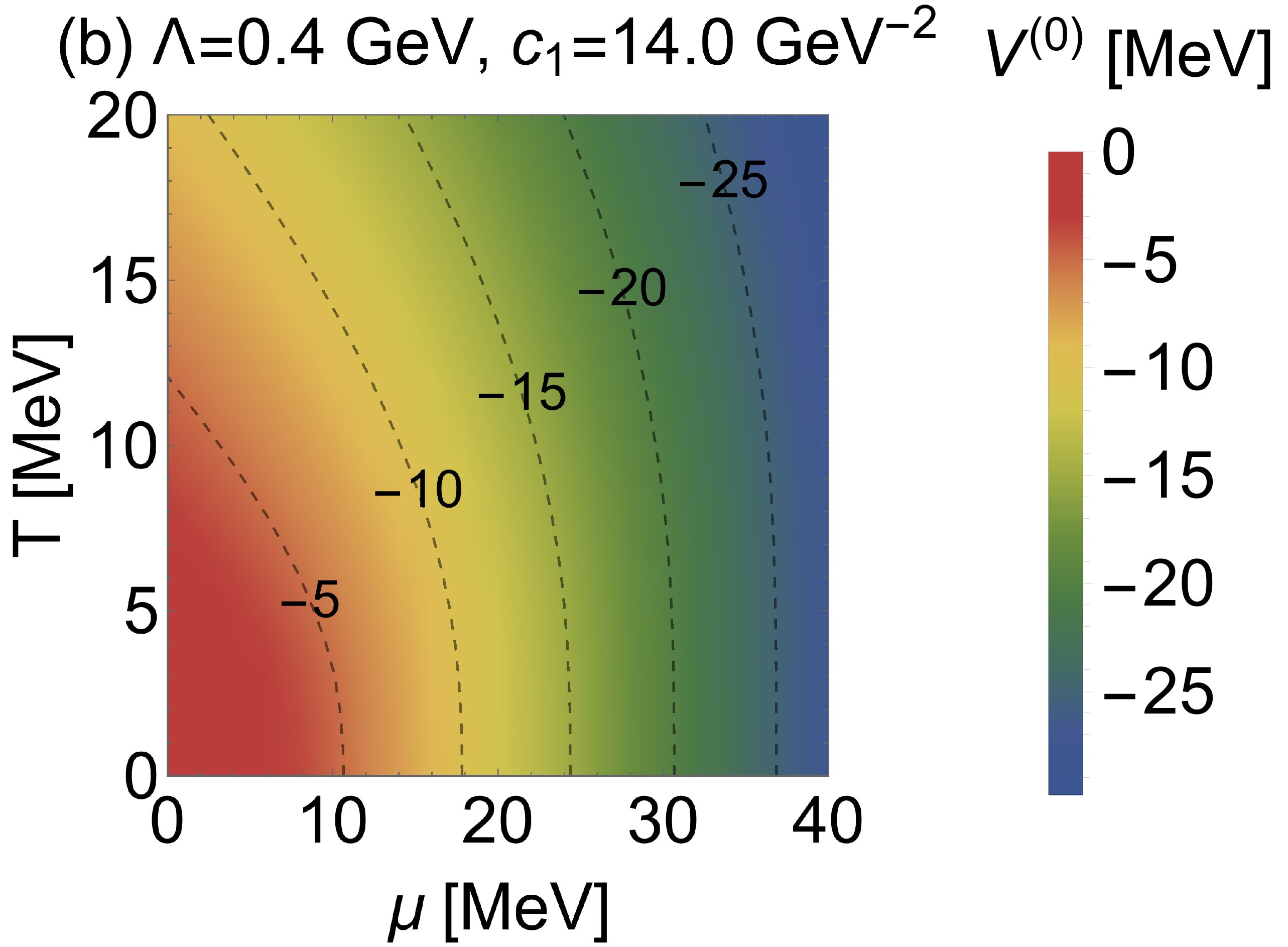}
\includegraphics[scale=0.22]{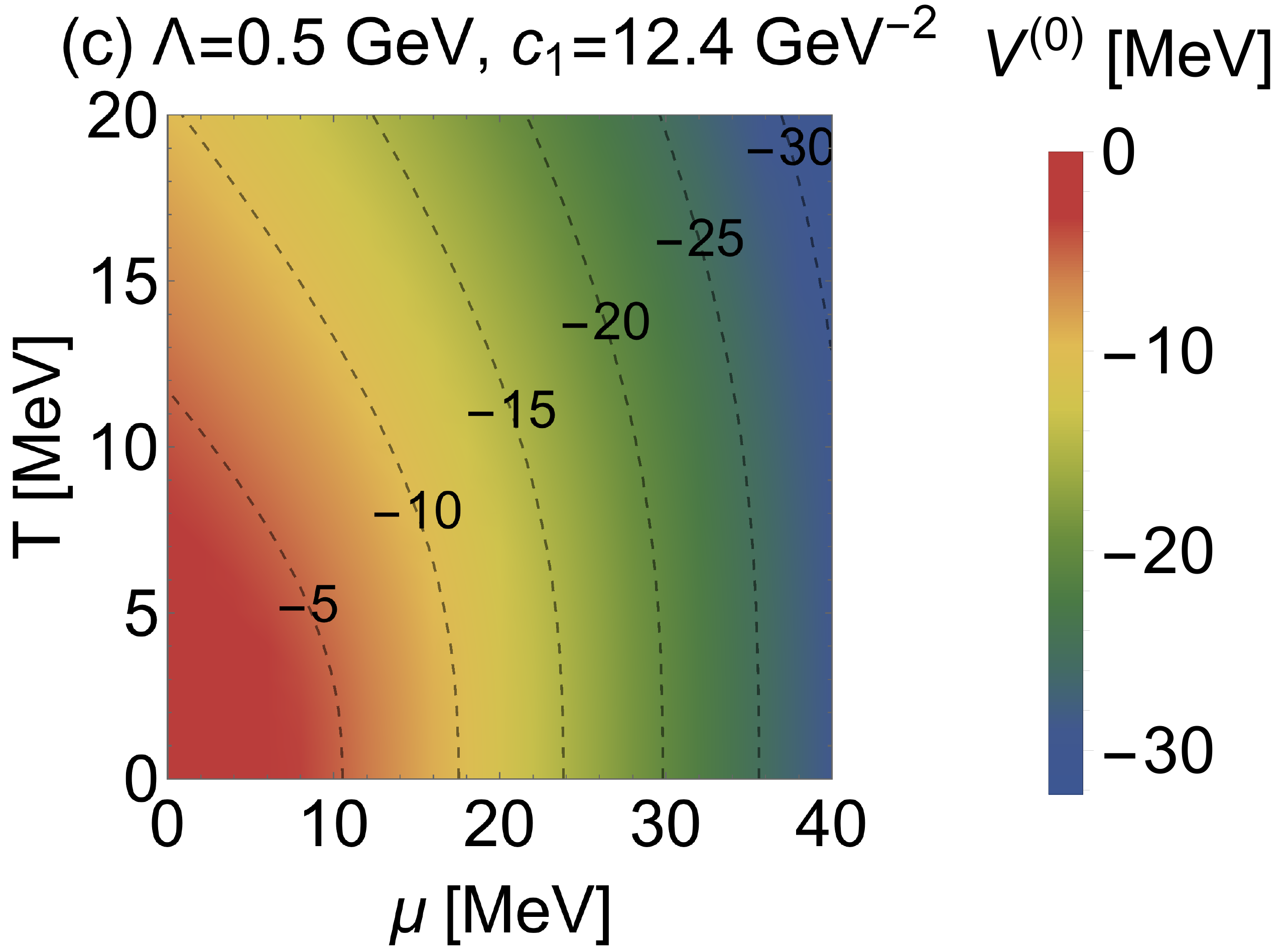}
\vspace{0em}
\caption{The effective potentials $V^{(0)}=V^{(0)}(T,\mu)$ are shown on the $\mu$-$T$ plane for the parameter sets (a), (b), and (c).}
\label{fig:Fig_V0_mu_T_FF_SC}
\end{center}
\end{figure}

\begin{figure}[tb]
\begin{center}
\vspace{0cm}
\includegraphics[scale=0.3]{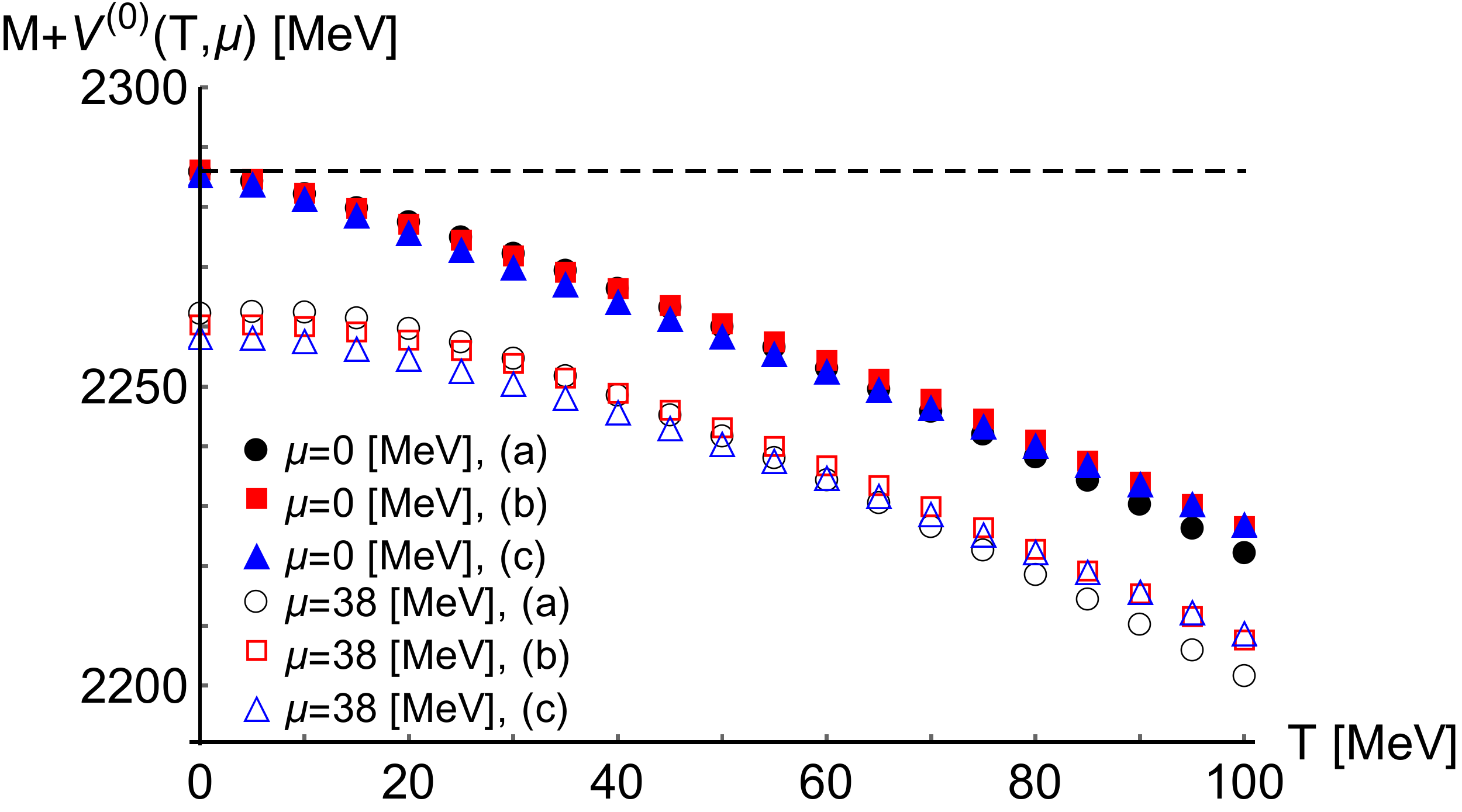}
\vspace{0em}
\caption{The mass of a $\Lambda_{c}$ baryon at rest in nuclear matter, $M+V^{(0)}(T,\mu)$ in Eq.~\eqref{eq:energy_0}, as functions of the temperature for the chemical potentials $\mu=0$ and $38$ MeV, respectively, for the parameter sets (a), (b), and (c). The horizontal dashed line indicates the mass of a $\Lambda_{c}$ baryon in vacuum. See also Fig.~\ref{fig:Fig_V0_mu_T_FF_SC}.}
\label{fig:190904}
\end{center}
\end{figure}

The values of the effective potential $V^{(0)}(T,\mu)$ at various temperature $T$ and chemical potential $\mu$ are shown on the $\mu$-$T$ plane in Fig.~\ref{fig:Fig_V0_mu_T_FF_SC}.
It is a reasonable result that the values of $V^{(0)}(T,\mu)$ become smaller, and hence that the binding energies become larger, as the chemical potential increases.
This is simply induced by the larger Fermi surface at larger chemical potential.
It is also found that the values of $V^{(0)}(T,\mu)$ become smaller as the temperature increases.
This result can be understood intuitively also, because the number density of nucleon gas increases as the temperature increases, and the probability for a nucleon to collide into the $\Lambda_{c}$ baryon should be enhanced.
I show explicitly the mass of $\Lambda_{c}$ baryon at rest in nuclear matter, $M+V^{(0)}(T,\mu)$ in Eq.~\eqref{eq:energy_0}, as functions of the temperature in Fig.~\ref{fig:190904}.

\subsection{Effective potential in moving frame and effective mass}
\label{sec:result_eff_mass}

\begin{figure}[tb]
\begin{center}
\vspace{0em}
\includegraphics[scale=0.4]{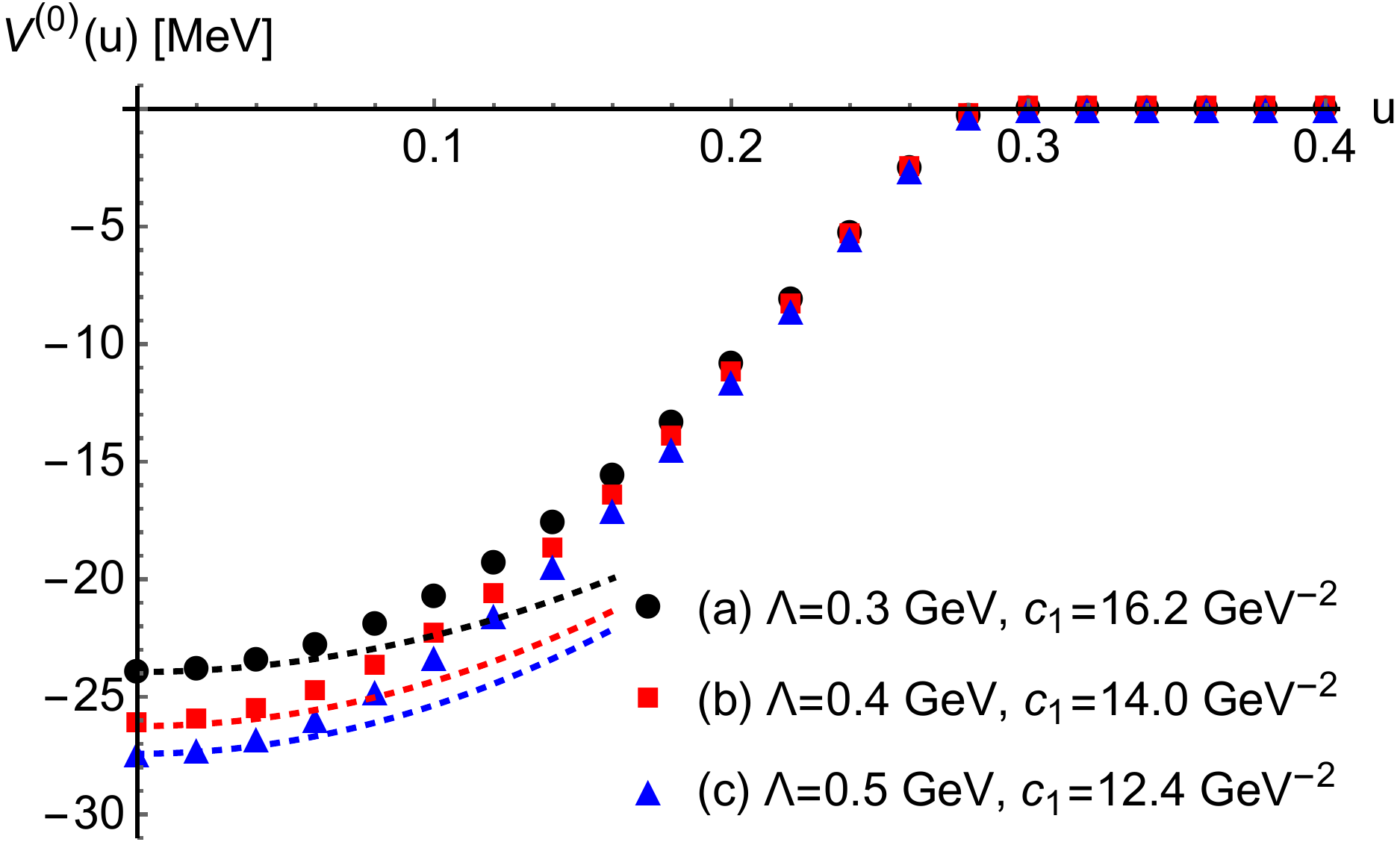}
\vspace{0em}
\caption{The plots of the effective potential $V^{(0)}(u)=V^{(0)}(T,\mu;\vec{u})$ as a function of three-dimensional velocity of the $\Lambda_{c}$ baryon $u=|\vec{u}|$ at $T=0$ MeV and $\mu=38$ MeV ($n_{N}=0.17$ fm$^{-3}$) for the parameter sets (a), (b), and (c). The dashed lines are the approximate curves for small $u$ in Eq.~\eqref{eq:eff_pot_u_expansion}.}
\label{fig:Fig_V0_u_mu_T_FF_SCabc}
\end{center}
\end{figure}

\begin{figure}[tb]
\begin{center}
\vspace{0cm}
\includegraphics[scale=0.22]{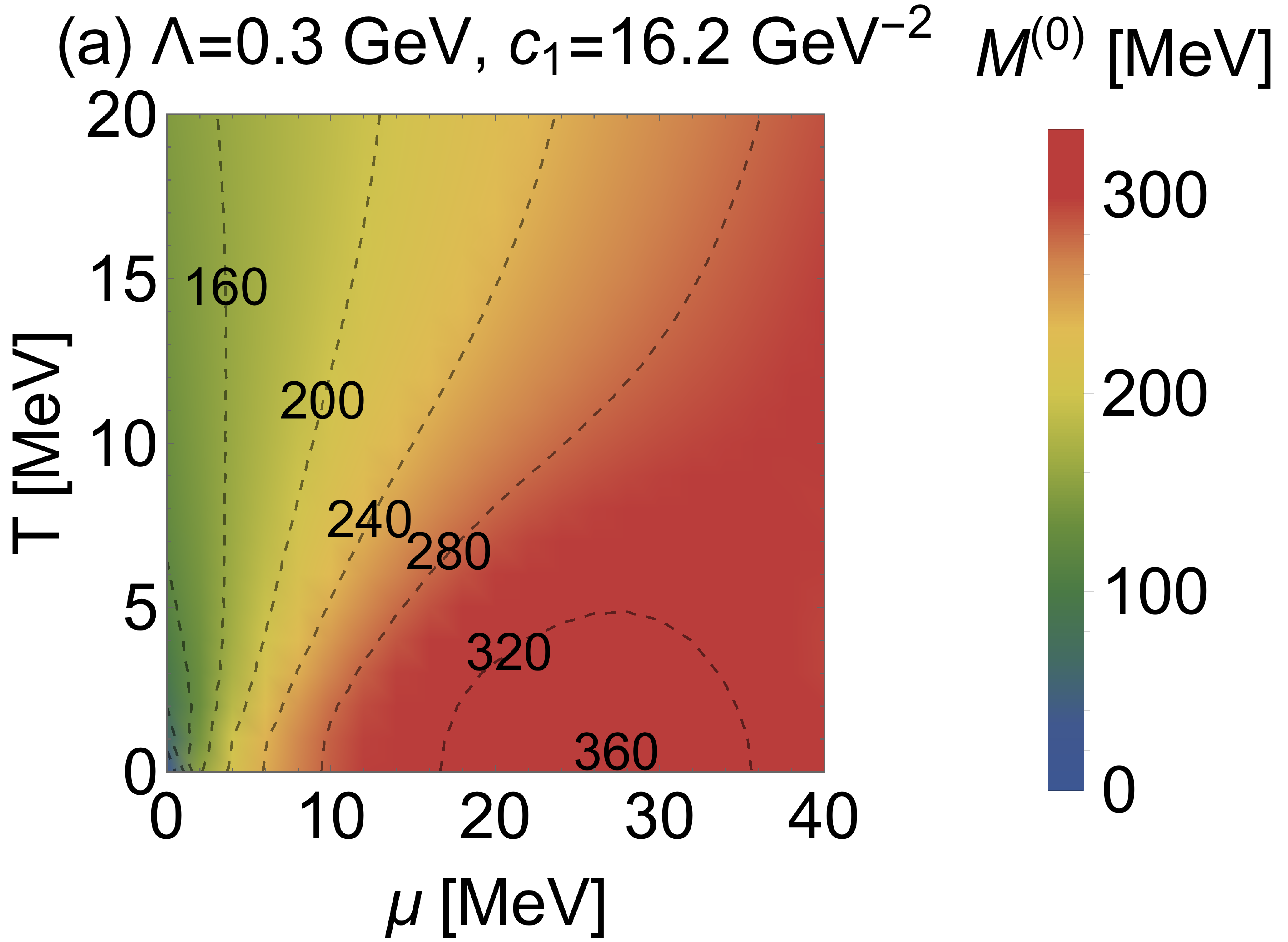}
\includegraphics[scale=0.22]{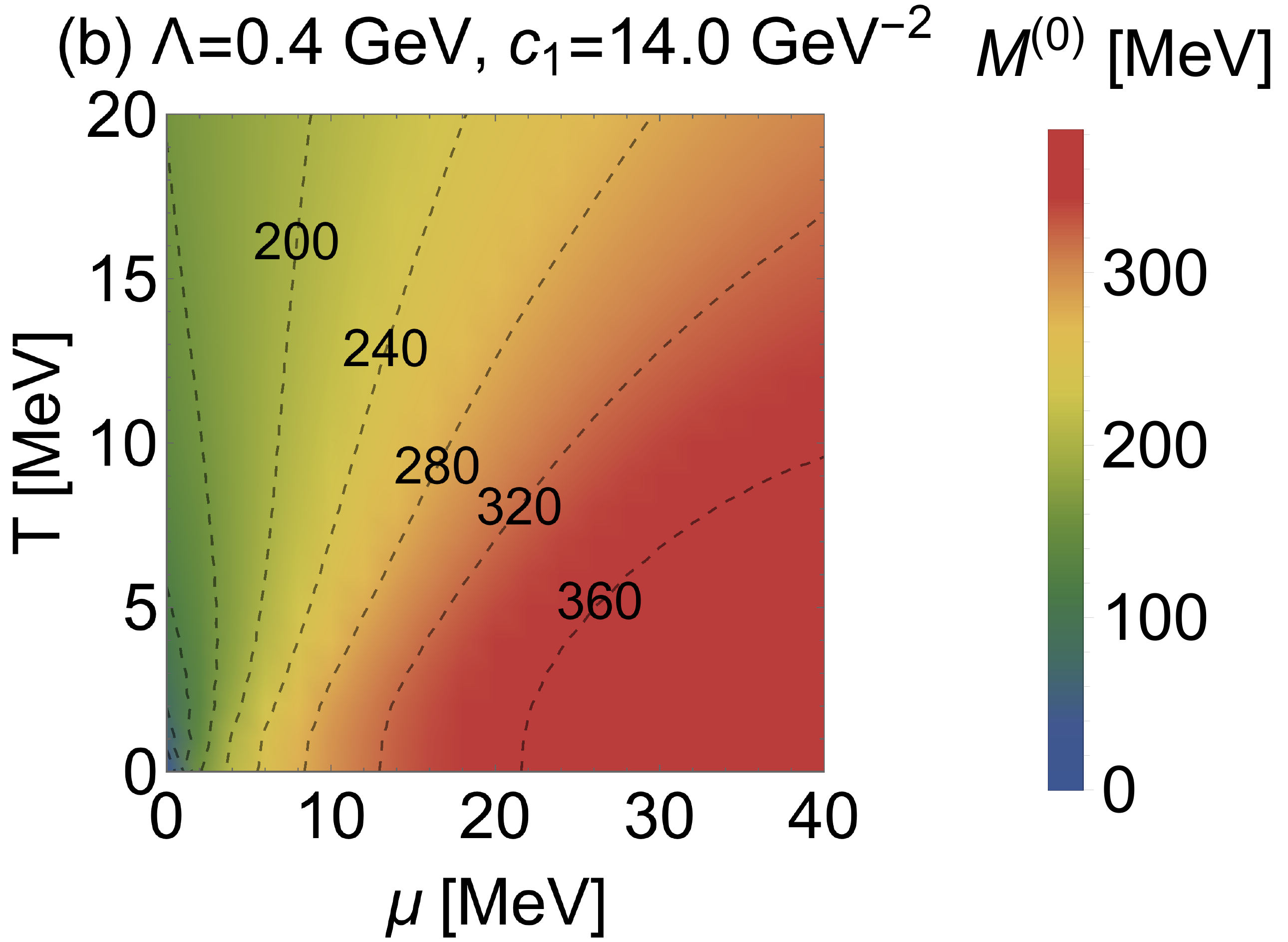}
\includegraphics[scale=0.22]{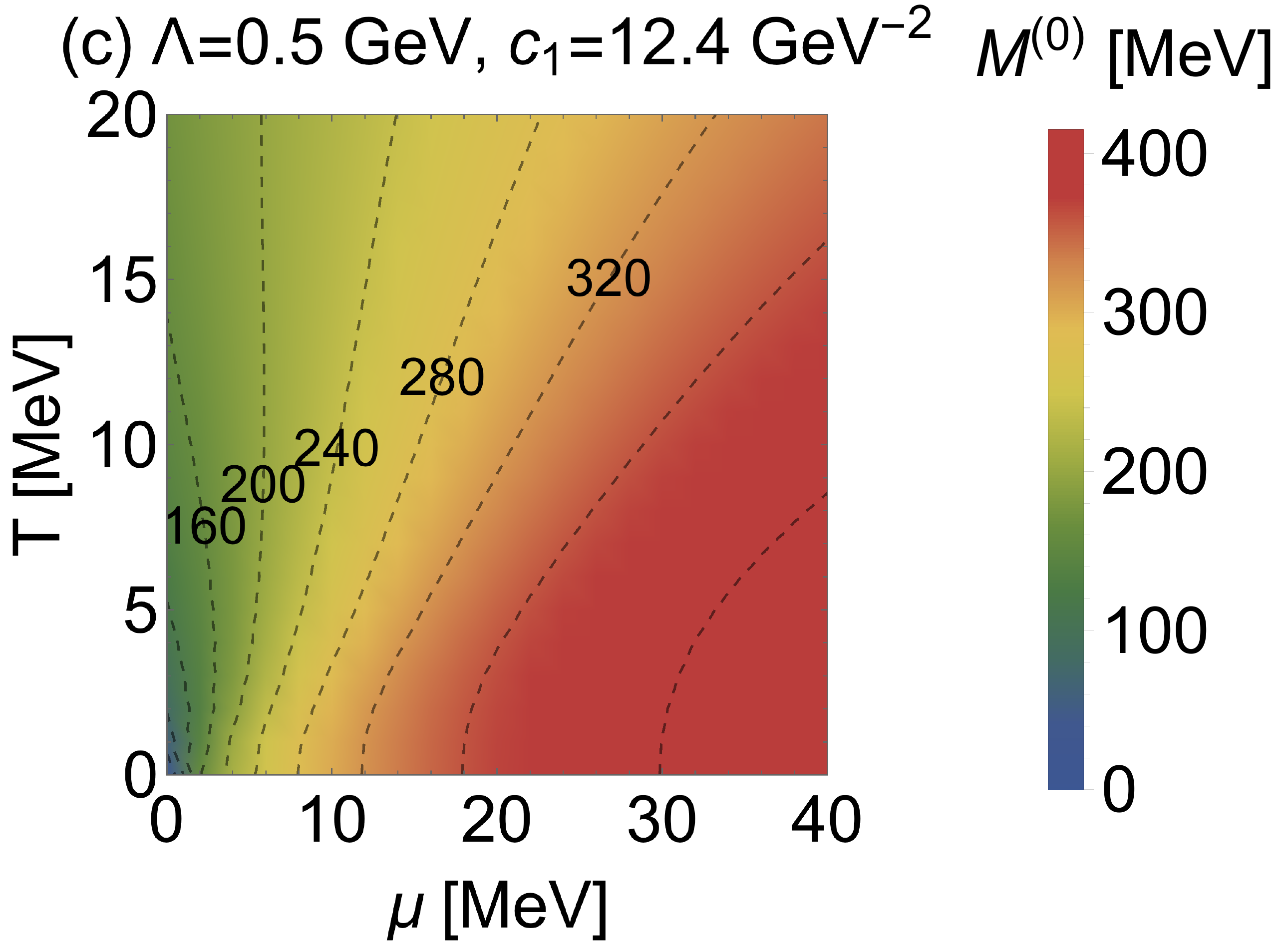}
\vspace{0em}
\caption{The effective masses $M^{(0)}=M^{(0)}(T,\mu)$ are shown on the $\mu$-$T$ plane for the parameter sets (a), (b), and (c).}
\label{fig:Fig_M0_mu_T_FF_SC}
\end{center}
\end{figure}

I plot the effective potentials at finite velocity $V^{(0)}(T,\mu;\vec{u})$, Eq.~\eqref{eq:eff_pot_u_T},  for the parameter sets (a), (b), and (c) in Fig.~\ref{fig:Fig_V0_u_mu_T_FF_SCabc}.
As the velocity $u=|\vec{u}|$ increases, the effective potentials becomes shallower and they eventually become zero at the critical velocity $u_{c}=0.28$.
It is seen that the approximate curves by Eq.~\eqref{eq:eff_pot_u_expansion} are appropriate for small $u=|\vec{u}|$.
The tendency that the effective potentials become shallower as the finite velocity increases can be understood in a naive manner, because the $\Lambda_{c}$ baryon moving in nuclear matter has a smaller probability to interact with nucleons.
The values of the effective mass $M^{(0)}(T,\mu)$ in Eq.~\eqref{eq:eff_mass_eff_pot_u_T} are calculated at zero temperature and normal nuclear-matter density.
The results are shown in Table~\ref{table:sharp_cutoff}.
They are in the range from $308$ to $415$ MeV in the present parameter sets.
I plot the results of the effective mass  $M^{(0)}(T,\mu)$ at various temperature and chemical potential on the $\mu$-$T$ plane in Fig.~\ref{fig:Fig_M0_mu_T_FF_SC}.
One finds the tendency that the effective masses increase at lager chemical potential for a fixed temperature, while they decrease at larger temperature for a fixed chemical potential.

\subsection{Change of number density of nucleons}

\begin{figure}[tb]
\begin{center}
\vspace{0cm}
\includegraphics[scale=0.4]{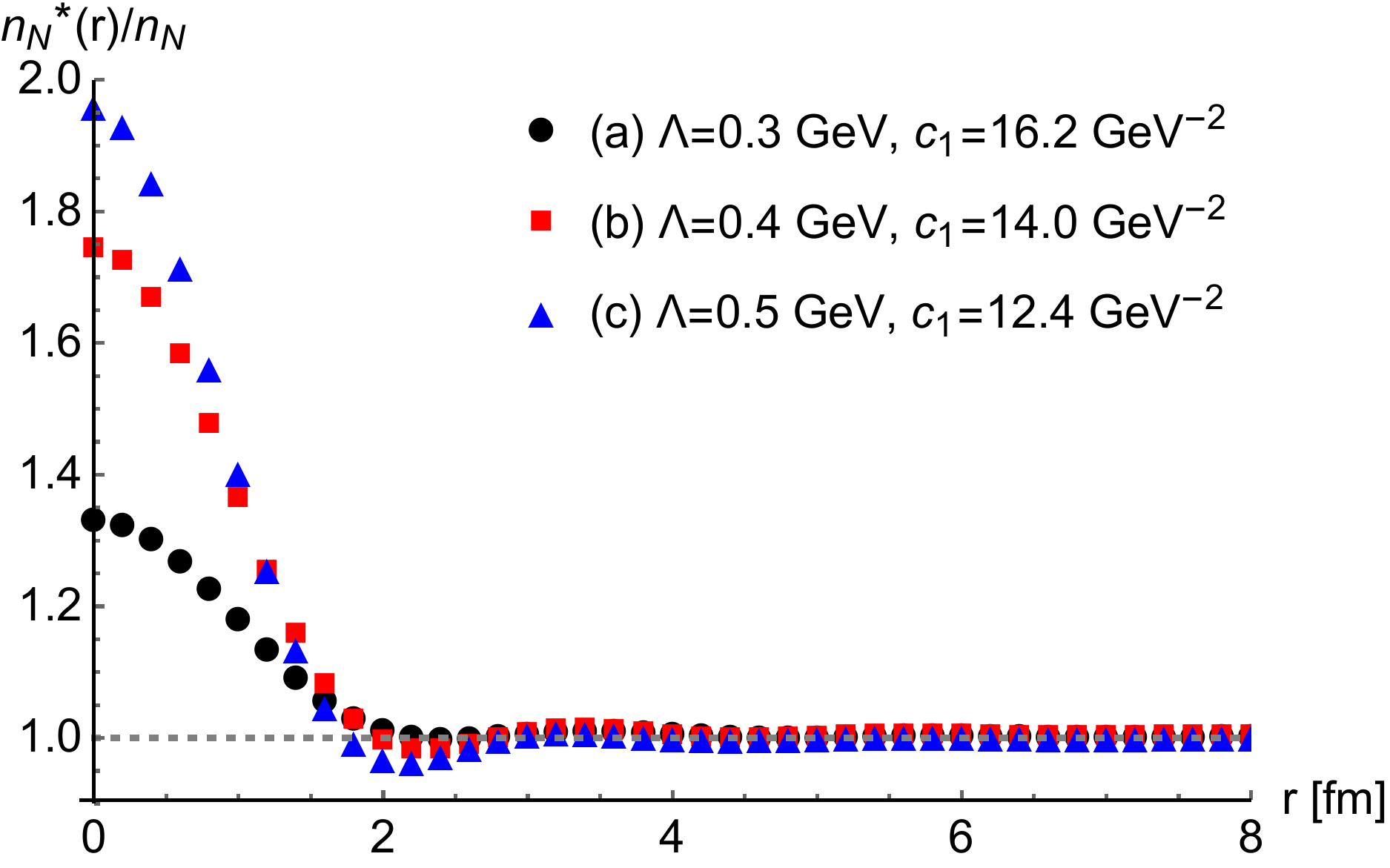}
\includegraphics[scale=0.25]{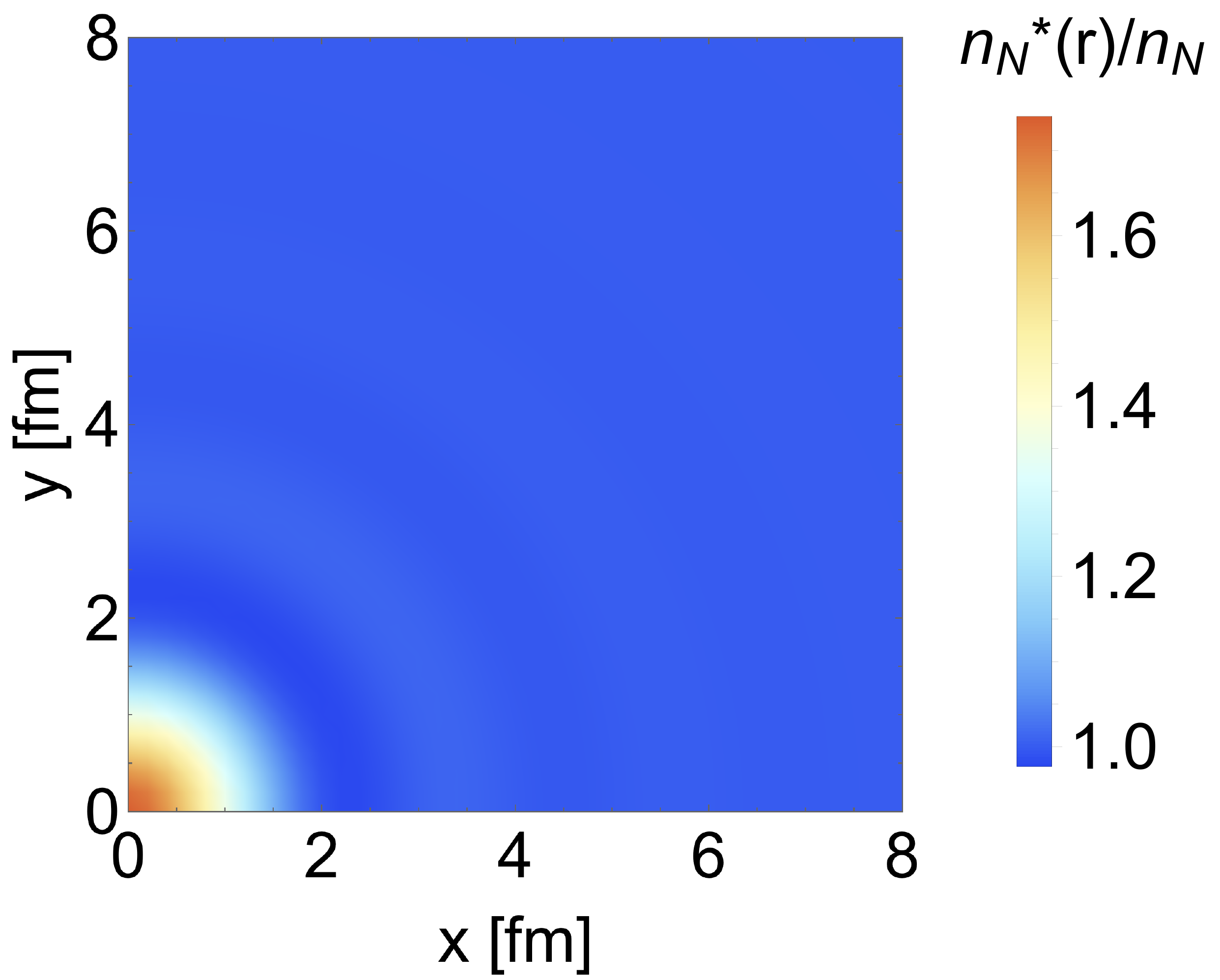}
\vspace{0em}
\caption{Left: The changes of nucleon number density $n_{N}^{\ast}(r)/n_{N}=n_{N}^{\ast}(T,\mu;\vec{x})/n_{N}(T,\mu)$ as a function of the distance $r=|\vec{x}|$ from the site of the $\Lambda_{c}$ baryon at $T=0$ MeV and $\mu=38$ MeV ($n_{N}=0.17$ fm$^{-3}$) are shown for the parameter sets (a), (b), and (c). Right: The same function for parameter set (b) is plotted on the $x$-$y$ plane with $r=\sqrt{x^{2}+y^{2}}$ as an intersection in space.}
\label{fig:Fig_n_r_T0_FF_SCabc}
\end{center}
\end{figure}

The change of nucleon number density $n_{N}^{\ast}(T,\mu;\vec{x})/n_{N}(T,\mu)$ is calculated by Eqs.~\eqref{eq:n_density_Lambdac} and \eqref{eq:delta_n_r}.
The values just at the $\Lambda_{c}$ baryon ($\vec{x}=\vec{0}$) are shown in Table~\ref{table:sharp_cutoff} [cf.~Eq.~\eqref{eq:delta_n_r0}].
They are in the range from 1.31 to 1.95 in the parameter sets (a), (b), and (c).
The enhancement is considered to be reasonable because nucleons should feel an attraction to the $\Lambda_{c}$ baryon due to the negative value of the effective potential, and they can gather around the $\Lambda_{c}$ baryon.
Therefore, the high-density state of nuclear matter can be realized around the $\Lambda_{c}$ baryon.
Thus, a $\Lambda_{c}$ baryon is a useful probe to study the higher-density state.
One notices that this high-density state exists just near the $\Lambda_{c}$ baryon, and it is reduced to the normal nucleon number density at far distances.
The spatial dependence is plotted in Fig.~\ref{fig:Fig_n_r_T0_FF_SCabc}.
The high-density state appears locally within the finite distance $r \simle 2$ fm around the position of the $\Lambda_{c}$ baryon.
One finds that the change of the nucleon number density damps with small oscillations at the distances $r \simge 2$ fm.
This is regarded as the Friedel oscillation which is known in the condensed-matter physics.

\section{Discussions}
\label{sec:discussions}

\subsection{Comparison to QCD sum rules}

Let us compare our results with the ones obtained by the analysis in the QCD sum rules.
The dispersion relation in Eq.~\eqref{eq:energy_Yasui} would be comparable with another form of dispersion relation of a $\Lambda_{c}$ baryon in nuclear matter,
\begin{eqnarray}
   E_{\Lambda_{c}}^{\ast(\mathrm{NM})}(T,\mu;\vec{q})
=
   \Sigma_{v}(T,\mu) + \sqrt{M^{\ast}(T,\mu)^{2}+\vec{q}^{2}},
\label{eq:energy_Ohtani0}
\end{eqnarray}
with the vector-type self-energy $\Sigma_{v}(T,\mu)$, the effective mass $M^{\ast}(T,\mu)$, and $\vec{q}$ the three-dimensional momentum of the $\Lambda_{c}$ baryon~\cite{Ohtani:2017wdc} (see also Ref.~\cite{Cohen:1994wm}).
The values of $\Sigma_{v}(T,\mu)$ and $M^{\ast}(T,\mu)$ were estimated in the QCD sum rules~\cite{Ohtani:2017wdc}.
Expanding Eq.~\eqref{eq:energy_Ohtani0} for small $\vec{q}$ and substituting $\vec{q}=M\vec{u}$ as the nonrelativistic form for small $\vec{u}$, one obtains
\begin{eqnarray}
   E_{\Lambda_{c}}^{\ast(\mathrm{NM})}(T,\mu;\vec{q})
&=&
   \Sigma_{v}(T,\mu) + M^{\ast}(T,\mu) + \frac{\vec{q}^{2}}{2M^{\ast}(T,\mu)} + {\cal O}(\vec{q}^{4})
\nonumber \\
&=&
   \Sigma_{v}(T,\mu) + M^{\ast}(T,\mu) + \frac{M^{2}}{2M^{\ast}(T,\mu)}\vec{u}^{2} + {\cal O}(\vec{u}^{4}).
\label{eq:energy_Ohtani}
\end{eqnarray}
Comparing Eq.~\eqref{eq:energy_Yasui} and Eq.~\eqref{eq:energy_Ohtani}, one finds
\begin{eqnarray}
   V^{(0)}(T,\mu) &=& \Sigma_{v}(T,\mu) + M^{\ast}(T,\mu)-M,
\label{eq:energy_Ohtani_V} \\
   M^{(0)}(T,\mu) &=& -\biggl( 1 - \frac{M}{M^{\ast}(T,\mu)} \biggr) M.
\label{eq:energy_Ohtani_M}
\end{eqnarray}
In the QCD sum rules, the values of $\Sigma_{v}(T,\mu)=-0.011$ GeV and $M^{\ast}(T,\mu)=2.277$ GeV were obtained at $T=0$ MeV and $\mu=38$ MeV ($n_{N}=0.17$ fm$^{-3}$)~\cite{Ohtani:2017wdc}.
Then, one has $\Sigma_{v}(T,\mu) + M^{\ast}(T,\mu)-M=-20$ MeV and $-\bigl( 1 - M/M^{\ast}(T,\mu) \bigr) M=9$ MeV in the right-hand sides in Eqs.~\eqref{eq:energy_Ohtani_V} and \eqref{eq:energy_Ohtani_M}, respectively.
The former is consistent with the value of the effective potential $V^{(0)}(T,\mu)$ obtained in Sec.~\ref{sec:result_eff_pot_rest}.
On the other hand, the latter is much smaller than the value of the effective mass $M^{(0)}(T,\mu)$ obtained in Sec.~\ref{sec:result_eff_mass}, though its sign is the same.
The difference may be due to the ambiguity at next-to-leading order ${\cal O}(1/M)$, because the terms at ${\cal O}(1/M)$ in the Lagrangian~\eqref{eq:Lagrangian_int} are not fully taken into account in the present study.
Furthermore, the higher-order terms in the (chiral) derivative expansion with higher momenta would be also important, because the terms with nonzero momenta would affect directly the effective mass at nonzero velocities. 
Those issues need to be addressed in the future.

\subsection{Diquarks in nuclear medium}

As discussed in the Introduction, the $ud$ diquark is an important subcomponent inside the $\Lambda_{c}$ baryon.
I suppose that the effective potential in the presence of the $\Lambda_{c}$ baryon in nuclear matter is the same as the mass shift of the $\Lambda_{c}$ baryon [cf.~Eq.~\eqref{eq:energy_0}] and consider that the mass shift is induced by the change of the constituent quarks inside the $\Lambda_{c}$ baryon.
Adopting the quark model, I suppose also that the baryon mass is given by a sum of the constituent quark masses and the energy from the spin-dependent interaction.
The Hamiltonian of the spin-dependent interaction between two quarks $i,j$ is expressed by
$
   H_{\mathrm{spin}}=(C_{B}/(m_{i}m_{j})) \vec{s}_{i} \!\cdot\! \vec{s}_{j}
$
where $C_{B}$ is the interaction constant, $m_{k}$ is the mass of the constituent quark $k=i,j$ and $\vec{s}_{k}$ is the spin operator acting on the constituent quark $k$.
I use $m_{q}=300$ MeV for $u$ and $d$ quarks, $m_{c}=1500$ MeV for the charm quark, and $m_{b}=4700$ MeV for the bottom quark and $C_{B}/m_{q}^{2}=193$ MeV to reproduce the mass splittings of normal hadrons in vacuum (see, e.g., Refs.~\cite{Lee:2007tn,Lee:2009rt}).
Here I consider not only the charm flavor but also the bottom flavor for generality of the discussion.

Inside the heavy baryons,
it is considered that there are the attractive $ud$ diquarks with spin 0 and isospin 0 in the $\Lambda_{Q}=\Lambda_{c}$, $\Lambda_{b}$ baryons, and the repulsive $ud$ diquarks with spin 1 and isospin 1 in the $\Sigma_{Q}=\Sigma_{c}$, $\Sigma_{b}$ and $\Sigma_{Q}^{\ast}=\Sigma_{c}^{\ast}$, $\Sigma_{b}^{\ast}$ baryons.
Those simple internal configurations are in good approximation as long as the heavy quark is sufficiently massive.
The mass of the $\Lambda_{Q}$ baryon can be parametrized by
\begin{eqnarray}
   M_{\Lambda_{Q}}(m_{q}) = m_{Q} + 2m_{q} - \frac{3}{4} \frac{C_{B}}{m_{q}^{2}} + c,
\label{eq:mass_Lambda_Q}
\end{eqnarray}
and the masses of the $\Sigma_{Q}$ and $\Sigma_{Q}^{\ast}$ baryons by
\begin{eqnarray}
   M_{\Sigma_{Q}}(m_{q}) = m_{Q} + 2m_{q} + \frac{1}{4} \frac{C_{B}}{m_{q}^{2}} - \frac{C_{B}}{m_{q}m_{Q}} + c,
\label{eq:mass_Sigma_Q_1} \\
   M_{\Sigma_{Q}^{\ast}}(m_{q}) = m_{Q} + 2m_{q} + \frac{1}{4} \frac{C_{B}}{m_{q}^{2}} + \frac{1}{2} \frac{C_{B}}{m_{q}m_{Q}} + c,
\label{eq:mass_Sigma_Q_2}
\end{eqnarray}
respectively, with $m_{Q}=m_{c}$, $m_{b}$.
In the above equations, $c$ is the energy constant stemming from the vacuum properties, such as color confinement, which are not included in the above model setups of the constituent quark and the diquark interaction.
In order to investigate the mass changes of the heavy baryons in nuclear matter, I consider that the light-quark mass $m_{q}$ is shifted to $m_{q}^{\ast}=m_{q}+\delta m_{q}$ in nuclear matter by partial restoration of the broken chiral symmetry.
Then, the heavy baryon masses in nuclear matter are given by
$ M_{\Lambda_{Q}}(m_{q}^{\ast}) = M_{\Lambda_{Q}}(m_{q}) + \delta M_{\Lambda_{Q}}$
for the $\Lambda_{Q}$ baryon and
$ M_{\Sigma_{Q}}(m_{q}^{\ast}) = M_{\Sigma_{Q}}(m_{q}) + \delta M_{\Sigma_{Q}}$ and $ M_{\Sigma_{Q}^{\ast}}(m_{q}^{\ast}) = M_{\Sigma_{Q}^{\ast}}(m_{q}) + \delta M_{\Sigma_{Q}^{\ast}}$
for the $\Sigma_{Q}$ and $\Sigma_{Q}^{\ast}$ baryons.
The result in Sec.~\ref{sec:numerical_results} indicates that the values $ M_{\Lambda_{Q}}(m_{q}^{\ast}) - M_{\Lambda_{Q}}(m_{q})$ are in the range from $-24.3$ MeV to $-28.1$ MeV at zero temperature and normal nuclear-matter density 
 (cf.~Table~\ref{table:sharp_cutoff}).
They give the mass shift $\delta m_{q}= -8$ MeV in average. 
Accordingly, the interaction energy between the $ud$ diquark, $- (3/4) (C_{B}/m_{q}^{2})=-144$ MeV in Eq.~\eqref{eq:mass_Lambda_Q}, is enhanced to $- (3/4) (C_{B}/m_{q}^{\ast 2})=-152$ MeV in the absolute value.
Thus the diquark becomes more bound in nuclear matter.
The above estimates may be too crude, but they will give us an interesting interpretation about the mass changes of the heavy baryons in nuclear matter.

With the value $\delta m_{q}= -8$ MeV, I obtain the mass shifts for the $\Sigma_{c}$ and $\Sigma_{c}^{\ast}$ baryons and for the $\Lambda_{b}$, $\Sigma_{b}$, and $\Sigma_{b}^{\ast}$ baryons at normal nuclear-matter density:
$\delta M_{\Sigma_{c}}=-15$ MeV and $\delta M_{\Sigma_{c}^{\ast}}=-14$ MeV for the charm baryons and $\delta M_{\Lambda_{b}}=-25$ MeV, $\delta M_{\Sigma_{b}}=-14$ MeV, and $\delta M_{\Sigma_{b}^{\ast}}=-13$ MeV for the bottom baryons.
Notice $\delta M_{\Lambda_{c}}=\delta M_{\Lambda_{b}}$ in the present framework because the heavy quark is decoupled from the $ud$ diquark as shown in Eq.~\eqref{eq:mass_Lambda_Q} and the heavy-flavor dependence of the heavy baryon mass is not included.

\section{Conclusion}
\label{sec:conclusion}

I have discussed the properties of the $\Lambda_{c}$ baryon in nuclear matter at zero or finite temperature.
Starting from the Lagrangian at the leading order in the $1/M$ expansion for the $\Lambda_{c}$ baryon mass $M$ and assuming that the $\Lambda_{c}$ baryon is at rest,
I have derived the effective Lagrangian for the $\Lambda_{c}$ baryon and the nucleons.
The parameters in the Lagrangian are constrained by the scattering length estimated in the chiral extrapolation from the lattice QCD simulations.
Adopting the one-loop calculation, I have obtained the effective potential which is the quantity measuring the binding energy of the $\Lambda_{c}$ baryon in nuclear matter.
I have extended the effective potential to the case when the $\Lambda_{c}$ baryon moves with a constant velocity.
I also have derived the change of the nucleon number density around the $\Lambda_{c}$ baryon in nuclear matter.

The numerical values of the effective potential indicate that the $\Lambda_{c}$ baryon can be bound with the binding energy of about 20 MeV at normal nuclear-matter density.
This value is consistent with the ones estimated in other theoretical approaches.
The binding energy becomes larger as the temperature and/or the nucleon number density increases.
The effective mass, i.e. the mass of inertia for the $\Lambda_{c}$ baryon moving in nuclear matter, is also obtained.
The nucleon number density near the $\Lambda_{c}$ baryon becomes higher than normal nuclear matter, and thus the $\Lambda_{c}$ baryon can be a useful probe to research the higher-density state.

As future prospects, it will be necessary to consider the higher-order terms in the $1/M$ expansion for the $\Lambda_{c}$ baryon, the finite range potential between the $\Lambda_{c}$ baryon and the nucleon, the interactions between nucleons, and so on.
Those effects can be analyzed by the path integral and the auxiliary field as shown in the present work.
The extension to $\Sigma_{c}$ and $\Sigma_{c}^{\ast}$ baryons is also interesting.
Because $\Sigma_{c}$ and $\Sigma_{c}^{\ast}$ baryons have a finite spin and a finite isospin,
it may be possible to study the phenomena related to the Kondo effect, which is an impurity effect caused by the non-Abelian (spin- and isospin-exchange) interaction between the impurity particle and the fermion gas~\cite{Yasui:2019ogk} (see also Refs.~\cite{Yasui:2013xr,Yasui:2016ngy,Yasui:2016hlz}).
Excited states of charm baryons such as $\Lambda_{c}^{\ast}(2595)$ and $\Lambda_{c}^{\ast}(2625)$ are also interesting objects, because they are related to the $D$ and $D^{\ast}$ dynamics in nuclear matter (see Refs.~\cite{Hosaka:2016ypm,Krein:2017usp} and the references therein).
Reaction mechanisms to produce charm baryon in atomic nuclei at experimental facilities should be studied further~\cite{Shyam:2016uxa,Shyam:2017gqp} (see also Refs.~\cite{Yamagata-Sekihara:2015ebw,Shyam:2016bzq}).
Those subjects are left for future work.

\section*{Acknowledgment}
The author thanks Makoto Oka, Keisuke Ohtani, and Takaya Miyamoto for fruitful discussions.
The author also thanks Tomokazu Miyamoto for valuable comments on the manuscript.
This work is supported by JSPS Grant-in-Aid for Scientific Research (KAKENHI Grant No.~25247036 and No.~17K05435) and by the Ministry of Education, Culture, Sports, Science (MEXT)-Supported Program for the Strategic Research Foundation at Private Universities ``Topological Science" (Grant No. S1511006). 

\bibliography{reference}

\end{document}